\documentclass{elsartNoFoot}

\usepackage{jb}
\usepackage[british]{babel}
\usepackage{epsf}
\usepackage{amssymb}
\usepackage{wasysym}	
\usepackage{url}
\usepackage{latin1}

\usepackage[pdftex]{color}

\usepackage{graphicx}

\usepackage[paperwidth=210mm,paperheight=297mm]{geometry}	
\definecolor{coLabel}	{rgb}{0.50,0.00,0.00}	
\definecolor{coRemark}	{rgb}{0.99,0.40,0.40}	

\definecolor{cWrong}	{rgb}{0.70,0.00,0.00}	
\definecolor{cRight}	{rgb}{0.00,0.70,0.00}	
\definecolor{cIstTerm}	{rgb}{0.50,0.20,0.00}	
\definecolor{cTestname} {rgb}{0.00,0.30,0.20}	

\definecolor{cWrongBg}	{rgb}{0.99,0.90,0.90}	
\definecolor{cRightBg}	{rgb}{0.90,0.99,0.90}	
\definecolor{cIrrelBg}	{rgb}{0.90,0.90,0.99}	
\definecolor{cIrrel}	{rgb}{0.00,0.00,0.70}	

\definecolor{coThmName}	{rgb}{0.00,0.00,0.99}
\definecolor{coNotion}	{rgb}{0.20,0.00,0.20}	

\definecolor{cFilename} {rgb}{0.00,0.00,0.50}
\definecolor{cProc}     {rgb}{0.00,0.20,0.30}

\definecolor{coIstTerm}	{rgb}{0.50,0.20,0.00}	
\definecolor{coTestname} {rgb}{0.00,0.30,0.20}	
\definecolor{coUnused}	{rgb}{0.70,0.70,0.70}	

\definecolor{coSeqPos}	{rgb}{0.30,0.30,0.99}	

\definecolor{coVal} 	{rgb}{0.00,0.00,0.50}	
\definecolor{coTerm}	{rgb}{0.00,0.50,0.50}	
\definecolor{coWgTerm}	{rgb}{0.50,0.00,0.00}	
\definecolor{coWg}	{rgb}{0.50,0.50,0.00}	
\definecolor{coWgSep}	{rgb}{0.50,0.50,0.50}	
\definecolor{coOpSep}	{rgb}{0.90,0.90,0.90}	

\definecolor{coHeapFg}	{rgb}{1.00,0.00,0.00}	
\definecolor{coHeapBg}	{rgb}{1.00,0.95,0.95}	
\definecolor{coHeapLn}	{rgb}{1.00,0.50,0.50}	
\definecolor{coDoneFg}	{rgb}{0.00,0.00,1.00}	
\definecolor{coDoneBg}	{rgb}{0.95,0.95,1.00}	
\definecolor{coDoneLn}	{rgb}{0.50,0.50,1.00}	

\definecolor{cConv} {rgb}{0.70, 0.70, 0.70}
\definecolor{cVar}  {rgb}{0.40, 0.00, 0.00}
\definecolor{cConst}{rgb}{0.40, 0.60, 0.40}

\definecolor{cIdxUnsel}	{rgb}{0.70,0.70,0.70}
\definecolor{cIdxSel}	{rgb}{0.00,0.00,0.00}

\definecolor{cMInc}	{rgb}{0.90,0.90,0.90}	
\definecolor{cIncl}	{rgb}{0.50,0.50,0.50}	

\definecolor{cT0}       {rgb}{0.00,0.00,0.00}
\definecolor{cT1}       {rgb}{0.55,0.55,0.55}

\definecolor{coEv}	{rgb}{0.99,0.60,0.60}
\definecolor{coEe}	{rgb}{0.00,0.60,0.00}
\definecolor{coTm}	{rgb}{0.00,0.00,0.60}

\newcommand{\LABEL}[1]{\label{#1}}

\newcommand{\Lra}{\Leftrightarrow}
\newcommand{\ra}{\rightarrow}
\newcommand{\raa}{\longrightarrow}
\newcommand{\Ra}{\Rightarrow}
\renewcommand{\leq}{\leqslant}
\renewcommand{\geq}{\geqslant}

\newcommand{\tpl}[1]{\langle #1 \rangle}	
\newcommand{\set}[1]{\{ #1 \}}			
\newcommand{\subst}[1]{\{ #1 \}}		
\newcommand{\sortdef}{::=}

\newcommand{\wg}[1]{\overline{#1}}	
\newcommand{\nf}{{\sf nf}}		

\newcommand{\eqc}[3]{\mbox{\bf [}#1\mbox{\bf ]}_{#2}^{#3}}

\newcommand{\V}{{\cal V}}		
\newcommand{\W}{{\cal W}}		
\renewcommand{\L}{{\cal L}}		
\newcommand{\G}{{\cal G}}		
\renewcommand{\O}{{\cal O}}		

\newcommand{\NF}{{\sf NF}}		
\newcommand{\F}{{\sf F}}		
\newcommand{\Fhist}{{\sf F}_{\sf hist}}	
\newcommand{\D}{{\sf D}}		

\newcommand{\N}{I\!\!N}			

\newcommand{\bigmid}{\rule[-0.07cm]{0.04cm}{0.4cm}\hspace*{0.1cm}}

\newcommand{\idiv}{\mathop{/\!/}}

\newcommand{\powerset}{\wp}

\newcommand{\T}[3]{%
	\Ite{#1}%
		{\Ite{#2}%
			{\Ite{#3}%
				{{\cal T}_{#3,#2}(#1)}
				{{\cal T}_{#2}(#1)}
			}%
			{\Ite{#3}%
				{{\cal T}_{#3}(#1)}
				{{\cal T}(#1)}
			}%
		}%
		{\Ite{#2}%
			{\Ite{#3}%
				{{\cal T}_{#3,#2}}
				{{\cal T}_{#2}}
			}%
			{\Ite{#3}%
				{{\cal T}_{#3}}
				{{\cal T}}
			}%
		}%
	}

\renewcommand{\:}[4]{%
        {%
        \renewcommand{\:}[4]{%
                {%
                \renewcommand{\:}[4]{error\error}%
                \renewcommand{\j}{{##2}}%
                {##4}%
                ##1...##1%
                \renewcommand{\j}{{##3}}%
                {##4}%
                }%
        }%
        \renewcommand{\i}{{#2}}%
        {#4}%
        #1\ldots#1%
        \renewcommand{\i}{{#3}}%
        {#4}%
        }%
}

\newcommand{\bigOp}[5]{%
	{%
	\renewcommand{\i}{#1}%
	\renewcommand{\bigOp}[5]{%
		{%
		\renewcommand{\j}{#1}%
		\renewcommand{\bigOp}[5]{error\error}%
		#2_{#1=#3}^{#4} #5%
		}%
	}%
	#2_{#1=#3}^{#4} #5%
	}%
}

\newcommand{\THM}[3]{%
	\begin{#1}%
	\Ite{#2}{ \textcolor{coThmName}{(#2)}\\ }{ }%
	{#3}%
	\end{#1}%
	}
\newcommand{\PRF}[2]{%
	\begin{pf}%
	{#2}%
	\end{pf}%
	}
\newcommand{\EXAMPLE}[2]   {\THM{example}   {#1}{#2}}
\newcommand{\LEMMA}[2]     {\THM{lemma}     {#1}{#2}}
\newcommand{\THEOREM}[2]   {\THM{theorem}   {#1}{#2}}
\newcommand{\DEFINITION}[2]{\THM{definition}{#1}{#2}}
\newcommand{\ALGORITHM}[2] {\THM{jbalgorithm} {#1}{#2}}

\newcommand{\PROOF}[1]     {\PRF{\em Proof. }       {#1}}

\newcommand{\cal}{\mathcal}

\renewcommand{\.}[1]{\!#1\!}

\newcommand{\notion}[1]{\emph{\textcolor{coNotion}{#1}}}

\newcommand{\true}{{\sf t}}
\newcommand{\false}{{\sf f}}
\newcommand{\undef}{{\sf u}}

\newcommand{\idx}{{\mbox{\sf idx}}}

\newcommand{\code}[1]{{\tt #1}}

\begin{document}

\newsavebox{\atpic}
\savebox{\atpic}{\includegraphics[scale=0.08]{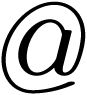}}

\setReceivedPrefix{}

\begin{frontmatter}

\title{An Improved Algorithm for E-Generalization}
\author{Jochen Burghardt}
\address{jochen.burghardt\usebox{\atpic}alumni.tu-berlin.de}
\received{Sep 2017}

\begin{abstract}
E-generalization computes common generalizations of given ground terms
w.r.t.\ a given equational background theory $E$.
~
In~\cite{Burghardt.2005c}, we had presented a computation
approach based on standard regular tree grammar algorithms,
and a {\sc Prolog} prototype implementation.
~
In this report, we present algorithmic improvements, prove them
correct and complete, and give some details
of an efficiency-oriented implementation in {\tt C} that allows us to
handle problems larger by several orders of magnitude.
\end{abstract}


\end{frontmatter}

\newpage
\tableofcontents


\newpage
\section{Introduction}
\LABEL{Introduction}

\notion{E-generalization} computes common abstractions of
given objects, considering given background
knowledge to obtain most suitable results.

More formally,
assuming familiarity with term rewriting theory 
(e.g.\ \cite{Dershowitz.Jouannaud.1990a}),
the problem of E-generalization consists in computing the set
of common generalizations of given ground terms w.r.t.\ a given
equational background theory $E$.
~
In~\cite[Def.1,2]{Burghardt.2005c}, we gave a formal definition of
the notion of a \notion{complete set of $E$-generalizations of given
ground terms}, and presented [Thm.7] an approach to compute such a
set by means of standard algorithms on regular tree grammars
(e.g.\ \cite{Thatcher.Wright.1968,Comon.Dauchet.Gilleron.2001}).
~
Its most fruitful application turned out to be the synthesis of
nonrecursive function definitions from input-output examples:
~
given some variables $\:,1n{x_\i}$, 
some ground substitutions $\:,1m{\sigma_\i}$, 
and some ground terms $\:,1m{t_\i}$,
we construct a term $t[\:,1n{x_\i}]$ such that
\[\begin{array}{ccc}
t[\:,1n{x_\i \sigma_1}] & =_E & t_1	\\
\vdots && \vdots	\\
t[\:,1n{x_\i \sigma_m}] & =_E & t_m	\\
\end{array}\]

\begin{figure}[b]
\begin{center}
\newcommand{\0}[1]{\textcolor{coSeqPos}{#1}}
$\begin{array}{r@{}c@{}c c l}
x & + & 0 & = & x	\\[-1ex]
x & + & s(y) & = & s(x+y)	\\[-1ex]
x & * & 0 & = & 0	\\[-1ex]
x & * & s(y) & = & x*y+x	\\[-1ex]
\end{array}$
\hfill
$\begin{array}{l|r|r||r|r|r||r|}
\cline{2-7}
\mbox{\0{Pos}} & ~\0{0} & ~\0{1} & ~\0{2} & ~\0{3} & ~\0{4}
	& \color{coUnused}~5	\\
\cline{2-7}
\mbox{Val} &  0 & 1 & 4 & 9 & 16 & \color{coUnused}25	\\
\cline{2-7}
\end{array}$
\hfill
$\begin{array}{c@{}c@{}c@{}c@{}c@{}c@{}c c r}
   & \0{v_p} &   & v_2 &   & v_1	\\	
\cline{2-6}
t[ & \0{2}   & , & 0   & , & 1   & ] & =_E & 4	\\
t[ & \0{3}   & , & 1   & , & 4   & ] & =_E & 9	\\
t[ & \0{4}   & , & 4   & , & 9   & ] & =_E & 16	\\
\end{array}$
\caption{Background theory (l), ~ sequence example (m), ~ sequence law synthesis (r)}
\LABEL{Background theory (l), sequence example (m), sequence law synthesis (r)}
\end{center}
\end{figure}

As a main example application of this approach,
we could obtain construction law terms for sequences.
~
For example, given the sequence $0, 1; 4, 9, 16$,
we could find a term $t[v_p,v_2,v_1]$
such that the equations in
Fig.~\ref{Background theory (l), sequence example (m), sequence law synthesis (r)}
(right) hold,
where $E$ consisted of the usual equations defining $(+)$ and $(*)$ on
natural numbers in $0$-$s$ notation (see
Fig.~\ref{Background theory (l), sequence example (m), sequence law synthesis (r)}
left),
writing e.g.\ $9$ for $s^9(0)$ for convenience.
~
Each such term $t[v_p,v_2,v_1]$ computes each supplied sequence member
$4$, $9$, and $16$ correctly from 
its 1st and 2nd predecessor $v_1$ and $v_2$,
respectively, and its position $v_p$ in the sequence (counting from $0$).
~
The supplied values $4$, $9$, $16$
are said to be \notion{explained} by the term
$t[v_p,v_2,v_1]$; in the sequence, we write a
\notion{semicolon} to separate an initial part that doesn't need to be
explained from an adjacent part that does.

For the given example, we obtained e.g.\ the
term $t[v_p,v_2,v_1] = v_p*v_p$;
when the background theory also included $(-)$,
we additionally obtained e.g.\
$t[v_p,v_2,v_1] = v_1+2*v_p-1$.
~
When instantiated by the substitution 
\mbox{$\subst{v_p \mapsto 5, v_2 \mapsto 9, v_1 \mapsto 16}$},
both terms yield
$25$ as sequence continuation,
which is usually accepted as a ``correct'' sequence extrapolation.
~
This way, $E$-generalization, or nonrecursive-function synthesis, can
be used to solve a certain class of intelligence tests.

In order to avoid solutions like $t[v_p,v_2,v_1] = (v_p+0*v_2)*v_p$,
we defined a notion of \notion{term weight} and focused on terms of
minimal weight for $t$.
~
As we computed a regular tree grammar describing all possible solution
terms in a compact form,
we could use a version of Knuth's algorithm from \cite{Knuth.1977} to
actually enumerate particular law terms in order of increasing weight.

$$*$$

The current report is in some sense the continuation
of~\cite{Burghardt.2005c}.
~
After giving some definitions in Sect.~\ref{Definitions},
it reports in
Sect.~\ref{An improved algorithm to compute E-generalizations}
algorithmic improvements on the problem of computing
$E$-generalizations.
~
Based on them, an efficiency-oriented reimplementation in {\tt C}
allows us to handle problems larger by
several orders of magnitude,
compared to the {\sc Prolog} implementation described in
\cite[Sect.~5.4]{Burghardt.2005c}.
~
Section~\ref{Implementation overview} discusses some details of that
implementation.
~
Section~\ref{Run time statistics} shows some run time statistics.

We used our implementation
to investigate the intelligence test \notion{IST'70-ZR}
\cite{Amthauer.1973};
a report on our findings is forthcoming.

\newpage
\section{Definitions}
\LABEL{Definitions}

\newcommand{\WG}{{\sf wg}}

\subsection{Terms}

\DEFINITION{Signature}{%
Let $\Sigma$ be
a finite set of function symbols, together with their
arities.
~
We abbreviate the set of all $n$-ary functions by $\Sigma_n$.
~
Let $\V$ be a finite set of variables.
\qed
}

\DEFINITION{Term}{%
For $V \subseteq \V$,
let $\T{}{}{V}$ denote the set of terms over $\Sigma$ with variables
from $V$, defined in the usual way.
~
In particular, $\T{}{}{\set{}}$ denotes the set of ground terms.
~
We use $(=)$ to denote the syntactic equality of terms.
}

\DEFINITION{Equality, normal form}{%
\LABEL{Equality, normal form}%
Let $E$ be a congruence relation on $\T{}{}{\set{}}$, i.e.\ an equivalence
relation that is compatible with each $f \in \Sigma$.
~
We require that $E$ is computable in the following sense:
~
a computable function $\nf: \T{}{}{\set{}} \raa \T{}{}{\set{}}$
shall exist such that
for all $t, t_1, t_2 \in \T{}{}{\set{}}$ we have
\begin{enumerate}
\item Idempotency: ~ $\nf(t) = \nf(\nf(t))$,
\item Decisiveness: ~ $t_1 =_E t_2 \;\Lra\; \nf(t_1) = \nf(t_2)$, and
\item Representativeness: ~ $t =_E \nf(t)$.
\end{enumerate}
In property~2, syntactic term equality is used on the right-hand side
of ``$\Lra$''.
~
Property 3 is redundant, it follows by applying 2 to 1.
~
We call $\nf(t)$ the \notion{normal form} of $t$, and say that $t$
\notion{evaluates} to
$\nf(t)$.
~
Let $\NF = \set{ \nf(t) \mid t \in \T{}{}{\set{}}}$
be the set of all normal
forms, also called \notion{values}.
\qed
}

Most often, $\nf$ is given by a confluent and terminating term
rewriting system (see e.g.\ \cite[p.245,267]{Dershowitz.Jouannaud.1990a});
the latter is obtained in turn most often from a conservative extension
of an initial algebra of ground constructor terms
(see e.g.\ \cite[Sect.7.3.2, p.160]{Duffy.1991}, \cite{Padawitz.1989a}).
~
As an example, let $\NF = \set{ s^i(0) \mid i \in \N}$ and let $\nf$
be given by the usual rewriting rules defining addition and
multiplication on that set, like those shown in 
Fig.~\ref{Background theory (l), sequence example (m), sequence law synthesis (r)} 
(left);
e.g.\ $\nf(s^2(0) * s^3(0)) = s^6(0) = \nf(s(0) + s^5(0))$.

\subsection{Weights}

In this subsection, we give the necessary definitions and properties
about term weights.
~
We associate to each function symbol $f$ a weight function $\wg{f}$ of
the same arity
(Def.~\ref{Weight function associated to a function symbol}) which
operates on some well-ordered domain $\W$ 
(Def.~\ref{Weight domain, minimal weight}).
~
Based on these functions, we inductively define the weight of a term
(Def.~\ref{Term weight, term set weight}).

\DEFINITION{Weight domain, minimal weight}{%
\LABEL{Weight domain, minimal weight}%
Let $\W$ be a set and $(<)$ an irreflexive total well-founded
ordering on $W$, such that $\W$ has a maximal element $\infty$.
~
We use $(\leq)$ to denote the reflexive closure of $(<)$.
~
Since $(<)$ is well-founded,
each non-empty subset $S$ of $\W$ contains a minimal
element $\min S \in S$;
we additionally define $\min \set{} := \infty$.
\qed
}

For our algorithm in
Sect.~\ref{An improved algorithm to compute E-generalizations},
we need a maximal element $\infty$ as initial value in
minimum-computation algorithms.
~
For the algorithm's completeness,
we additionally need the absence of limit ordinals
(e.g.\ \cite[Sect.19]{Halmos.1968})
below $\infty$ (see
Lem.~\ref{Properties of Alg. Weight decomposition list generation}).
~
Altogether,
for Alg.~\ref{Value-pair cached term generation}
we have to chose $\W$ order-isomorphic to
$\N \cup \set{\infty}$.
~
However,
the slightly more general setting from
Def.~\ref{Weight domain, minimal weight}
is more convenient in theoretical discussions and counter-examples.
~
We need $(<)$ to be total and well-founded in any case,
in order for a set of terms to contain a minimal-weight term.

\DEFINITION{Function properties wrt.\ order}{%
A function $\wg{f}: \W^n \raa \W$ is called
\begin{itemize}
\item
	increasing if
        $\bigwedge_{i=1}^n \; x_i \leq \wg{f}( \:,1n{x_\i} )$,
\item
	strictly increasing if
        $\bigwedge_{i=1}^n \; 
	( x_i < \infty \Ra x_i < \wg{f}( \:,1n{x_\i} ) )$,
\item
        monotonic if
        $(\bigwedge_{i=1}^n x_i \leq y_i)
        \Ra \wg{f}( \:,1n{x_\i} ) \leq \wg{f}( \:,1n{y_\i} )$,
\item
        strictly monotonic if
	it is monotonic and
	\\
        $(\bigwedge_{i=1}^n x_i < \infty) \wedge
        (\bigwedge_{i=1}^n x_i \leq y_i) \wedge
	(\bigvee_{i=1}^n x_i < y_i)
        \Ra \wg{f}( \:,1n{x_\i} ) < \wg{f}( \:,1n{y_\i} )$,
\item
	strict if
	$( \bigwedge_{i=1}^n w_i < \infty )
	\; \Ra \; \wg{f}(\:,1n{w_\i}) < \infty$,
	in particular, a $0$-ary weight function $\wg{f}$ is strict if
	$\wg{f}< \infty$.
\qed
\end{itemize}
}

Knuth's algorithm
---~which will play an important role below~---
requires weight functions to be monotonic and increasing
\cite[p.1l]{Knuth.1977}.

\DEFINITION{Weight function associated to a function symbol}{%
\LABEL{Weight function associated to a function symbol}%
Let $\W$ and $(<)$ as in Def.~\ref{Weight domain, minimal weight}.
~
Let a signature $\Sigma$ be given.
~
For each $n \in \N$ and $f \in \Sigma_n$, let a monotonic
and strictly increasing
weight function $\wg{f}: \W^n \raa \W$ be given;
we call $\wg{f}$ the weight function associated with $f$.
~
We assume that $\wg{f}(\:,1n{x_\i})$ can always
be computed in time $\O(n)$.
\qed
}

\DEFINITION{Term weight, term set weight}{%
\LABEL{Term weight, term set weight}%
Given the weight functions, define the weight $\WG(t)$
of a ground term $t$
inductively by
\[
\WG(f( \:,1n{t_\i} )) := \wg{f}( \:,1n{\WG(t_\i)} ) .
\]
For a set of ground terms $T \subseteq \T{}{}{\set{}}$,
define
\[
\WG(T) := \min \set{ \WG(t) \mid t \in T }
\]
to be the minimal weight of all terms in $T$.
~
Note that
$\WG(T) \in \W$ is always well-defined, and
for nonempty $T$ we have
$\WG(T) = \WG(t)$ for some $t \in T$,
since $\W$ is well-ordered.
\qed
}

\newcommand{\sz}{{\sf sz}}
\newcommand{\hg}{{\sf hg}}

\EXAMPLE{Weight function examples}{%
\LABEL{Weight function examples}%
The most familiar examples of weight measures are
the size $\sz(t)$, and the height $\hg(t)$ of a term $t$,
i.e.\ the total number of nodes,
and the length of the longest path from the root to any leaf,
respectively.
~
If $\W = \N \cup \set{ \infty }$
and $\wg{f}( \:,1n{x_\i} ) = 1 + \:+1n{x_\i}$
for each $f \in \Sigma_n$,
we get $\WG(t) = \sz(t)$;
the definitions $\wg{f}( \:,1n{x_\i} ) = 1 + \max \set{ \:,1n{x_\i} }$
for $f \in \Sigma_n$
yield $\WG(t) = \hg(t)$.
\qed
}

\EXAMPLE{Weight function counter-examples}{%
\LABEL{Weight function counter-examples}%
We give two counter-examples to show what our weight functions cannot
achieve.

First, it would be desirable in some contexts to prefer terms with
minimal sets of distinct variables.
~
Choosing $\W = \powerset(\V)$,
$\wg{v} = \set{v}$ for $v \in \V$,
and\footnote{
	relaxing $\wg{f}$'s increasingness from strict to nonstrict,
	for sake of the example
}
$\wg{f}( \:,1n{x_\i} ) = \:{\cup}1n{x_\i}$
for $f$ an $n$-ary function symbol (including constants),
we obtain as $\WG(t)$ the set of distinct variables occurring in the
term $t$.
~
However, we cannot use $\subsetneq$ as irreflexive
well-ordering on $\powerset(\V)$
in the sense of Def.~\ref{Weight domain, minimal weight},
since it is not total.
~
Even if in a generalized setting $<$ on $\W$ was allowed to be
a partial ordering, and we were interested only in the {\em number\/}
of distinct variables,
Knuth's algorithm from \cite{Knuth.1977} to compute minimal terms
cannot be generalized
accordingly to this setting, unless $P = NP$, as shown in
\cite[Sect.5, Lem.29]{Burghardt.2003a}.

Similarly, it can be desirable to have as $\WG(t)$ the number of distinct
subterms of $t$.
~
Following similar proof ideas as in \cite[Sect.5]{Burghardt.2003a},
it can be shown that this is impossible
unless the monotonicity requirement is dropped for weight functions,
and that Knuth's algorithm cannot be adapted to that setting,
again unless $P = NP$.
\qed
}

The following property will be used in the completeness and
correctness proof of Alg.~\ref{Value-pair cached term generation}
below.

\LEMMA{Strict weight-functions}{%
\LABEL{Strict weight-functions}%
If all weight functions are strict,
then $\forall t \in \T{}{}{\V}: \; \WG(t) < \infty$.
}
\PROOF{
Induction on the height of $t$.
\qed
}

\newcommand{\nt}[1]{N_{#1}}

\newpage
\section{An improved algorithm to compute E-generalizations}
\LABEL{An improved algorithm to compute E-generalizations}

In this section, we discuss some
algorithmic improvements on the problem of computing E-generalizations.
~
We first give the improved algorithm in Sect.~\ref{The algorithm}, and
prove its completeness and correctness in
Sect.~\ref{Correctness and completeness}
(see 
Thm.~\ref{Correctness and completeness of value-pair cached term generation}).
~
In Sect.~\ref{Relation to the grammar-based algorithm},
we relate it to the grammar-based algorithm from
\cite[Sect.~3.1]{Burghardt.2005c}, indicating that the former is an
improvement of the latter.
~
Implementation details are discussed in 
Sect.~\ref{Implementation overview}.

\subsection{The algorithm}
\LABEL{The algorithm}

If $f$ is an $n$-ary function symbol, and $w, \:,1n{w_\i} \in \W$ are
weights such that $\wg{f}(\:,1n{w_\i}) = w$,
we call the $n+1$-tuple $\tpl{f,\:,1n{w_\i}}$ a weight
decomposition list evaluating to $w$.

Our algorithm (Fig.~\ref{Example state in Alg. (red) and  (blue)}
shows an example state) can be decomposed in two layers.
~
The lower one
(Alg.~\ref{Weight decomposition list generation}, 
see left and red part in
Fig.~\ref{Example state in Alg. (red) and  (blue)}) generates, in order
of increasing evaluation weight, a sequence of
all possible weight decomposition lists.
~
The latter is fed into the higher layer
(Alg.~\ref{Term generation from decomposition lists},
see right and blue part in 
Fig.~\ref{Example state in Alg. (red) and  (blue)}), where each
decomposition list is used to generate a corresponding set of terms.

This pipeline
architecture is similar to that of a common compiler front-end,
where a scanner generates a sequence of tokens which is processed by
a parser.

\ALGORITHM{Weight decomposition list generation}{%
\LABEL{Weight decomposition list generation}%

Input:
\begin{itemize}
\item a signature $\Sigma$,
\item a computable weight function $\wg{f}$,
	for each function symbol $f \in \Sigma$
\item a finite set $V \subset \V$ of variables to be used in terms
\end{itemize}

Output:
\begin{itemize}
\item a potentially infinite stream of weight decomposition lists,
	with their evaluated weights being a non-decreasing sequence
\end{itemize}

Algorithm:
\begin{enumerate}
\item
	Maintain a set $\F$ of weight decomposition lists
	to be considered next.
\item
	Maintain a set $\Fhist$ of evaluating weights of
	all decomposition lists that have ever been drawn from $\F$.
\item\LABEL{Weight-gen heap init}
	Initially, let
	$\F = \set{ \tpl{v} \mid v \in V}
	\cup \set{ \tpl{c} \mid c \in \Sigma_0 }$
	be obtained from all variables and signature constants.
\item
	Initially, let $\Fhist = \set{}$.
\item\LABEL{Weight-gen empty}
	While $\F$ is non-empty, loop:
	\begin{enumerate}
	\item\LABEL{Weight-gen draw}
		Remove from $\F$ a weight decomposition list
		$\tpl{f, \:,1n{w_\i}}$
		evaluating to the least weight among all lists in $\F$;
		let $w$ denote that weight.
	\item
		Output $\tpl{f, \:,1n{w_\i}}$ to the stream.
	\item
		Insert $w$ into $\Fhist$.
	\item\LABEL{Weight-gen heap insert}
		If in step~\ref{Weight-gen draw}
		the least evaluating weight in $\F$ had increased since the
		previous visit,
		enter each possible weight decomposition list into
		$\F$ that can be built from $w$ and some weights
		from $\Fhist$.
		\\
		More formally: for every (non-constant) function symbol
		$f$ from the signature,
		enter into $\F$ each weight decomposition list
		$\tpl{f, \:,1n{x_\i}}$ such that
		$w \in \set{\:,1n{x_\i}} \subseteq \Fhist$.
	\qed
	\end{enumerate}
\end{enumerate}
}

\ALGORITHM{Term generation from decomposition lists}{%
\LABEL{Term generation from decomposition lists}%

Input:
\begin{itemize}
\item a pair of goal values $\tpl{g_1,g_2} \in \NF \times \NF$,
\item ground substitutions $\sigma_1, \sigma_2$,
\item a finite, or potentially infinite, stream of weight decomposition
	lists, ordered by ascending evaluation weight.
\end{itemize}

Output (if the algorithm terminates):
\begin{itemize}
\item a term $t$ of minimal weight
	such that $t \sigma_i =_E g_i$ for $i=1,2$.
\end{itemize}

Algorithm:
\begin{enumerate}
\item
	Maintain a partial mapping
	$\phi: \NF \times \NF \hookrightarrow \T{}{}{\V}$
	that yields the term of least weight considered so far (if any)
	for each value pair $v_1,v_2$
	such that $\phi(v_1,v_2) \sigma_1 =_E v_1$
	and $\phi(v_1,v_2) \sigma_2 =_E v_2$,
	if $\phi(v_1,v_2)$ is defined.
\item
	Maintain a set of
	minimal terms generated so far.
	~
	Terms will be added to it in order of increasing weight;
	so we can easily maintain a weight layer structure within it.
	~
	More formally:
	for each weight $w \in \W$,
	let $\D_w$ be the set of all minimal terms generated so far
	that have weight $w$.
\item\LABEL{cached term-gen init}
	Initially, let $\phi$ be the empty mapping.
\item
	Initially, let $\D_w = \set{}$, for each $w \in \W$.
\item\LABEL{cached term-gen empty}
	While the input stream of weight decomposition lists is non-empty,
	loop:
	\begin{enumerate}
	\item\LABEL{cached term-gen draw}
		Let $\tpl{f, \:,1n{w_\i}}$ be the next
		list from the stream,
		let $w$ denote the weight it evaluates to.
	\item\LABEL{cached term-gen loop over terms}
		For all $\:,1n{t_\i \in \D_{w_\i}}$, loop:
		\begin{enumerate}
		\item\LABEL{cached term-gen build}
			Build the term $t = f(\:,1n{t_\i})$.
			This term has weight $w$ by construction.
		\item\LABEL{cached term-gen eval}
			Let $v_1 = \nf(t \sigma_1)$,
			and $v_2 = \nf(t \sigma_2)$.
		\item\LABEL{cached term-gen new}
			If $\phi(v_1,v_2)$ is undefined,
			then
			\begin{enumerate}
			\item\LABEL{cached term-gen def phi}
				Add $\tpl{v_1,v_2} \mapsto t$ to $\phi$.
			\item
				Add $t$ to $\D_w$.
			\item\LABEL{cached term-gen goal}
				If $\tpl{v_1,v_2} = \tpl{g_1,g_2}$,
				then stop with success:
				\\
				$t$ is a term of minimal weight such
				that $t \sigma_i =_E g_i$.
			\end{enumerate}
		\end{enumerate}
	\end{enumerate}
\item\LABEL{cached term-gen fail}
	Stop with failure:
	no term $t$ with $t \sigma_i =_E g_i$ exists.
	\qed
\end{enumerate}
}

Note that the loop in step~\ref{cached term-gen empty} may continue
forever, if the input stream is infinite but no solution exists.
~
Next, we compose Alg.~\ref{Weight decomposition list generation}
and Alg.~\ref{Term generation from decomposition lists}:

\ALGORITHM{Value-pair cached term generation}{%
\LABEL{Value-pair cached term generation}%
Input:
\begin{itemize}
\item a signature $\Sigma$,
\item a computable weight function $\wg{f}$,
	for each function symbol $f \in \Sigma$
\item a pair of goal values $\tpl{g_1,g_2} \in \NF \times \NF$,
\item substitutions $\sigma_1, \sigma_2$.
\end{itemize}

Output:
\begin{itemize}
\item a term $t$ of minimal weight
	such that $t \sigma_i =_E g_i$ for $i=1,2$.
\end{itemize}

Algorithm:
\begin{itemize}
\item Let $V = \dom \sigma_1 \cup \dom \sigma_2$.
\item Feed $\Sigma$, all $\wg{f}$, and $V$
	into Alg.~\ref{Weight decomposition list generation}.
\item Feed $\tpl{g_1,g_2}$, $\sigma_1, \sigma_2$, and
	Alg.~\ref{Weight decomposition list generation}'s output
	stream into Alg.~\ref{Term generation from decomposition lists}.
\item Run Alg.~\ref{Weight decomposition list generation}
	in parallel to
	Alg.~\ref{Term generation from decomposition lists}
	until either algorithm terminates.
\end{itemize}

Implementation issues:
\begin{itemize}
\item The set $\F$ in
	Alg.~\ref{Weight decomposition list generation}
	is best realized by a heap
	(e.g.\ \cite[Sect.~3.4]{Aho.Hopcroft.Ullman.1974}).
\item Since we have
	$w \in \Fhist$
	in Alg.~\ref{Weight decomposition list generation}
	iff $\D_w \neq \set{}$
	in Alg.~\ref{Term generation from decomposition lists},
	the former test can be implemented by the latter one,
	thus avoiding the need for an implementation of $\Fhist$.
\item The mapping $\phi$ in
	Alg.~\ref{Term generation from decomposition lists}
	is best implemented by a hash table
	(e.g.\ \cite[Sect.~4.2]{Aho.Hopcroft.Ullman.1974})
	of balanced trees
	(e.g.\ \cite[Sect.~4.9]{Aho.Hopcroft.Ullman.1974}).
\item The sets $\D_w$ in
	Alg.~\ref{Term generation from decomposition lists}
	are just segments of one global list of terms,
	ordered by non-decreasing weight $w$.
\item The test $\tpl{v_1,v_2} = \tpl{g_1,g_2}$ in
	step~\ref{cached term-gen goal}
	of Alg.~\ref{Term generation from decomposition lists}
	can be speeded-up by initializing $\phi(g_1,g_2)$ to a special
	non-term entry ``${\sf goal}$''.
\qed
\end{itemize}
}

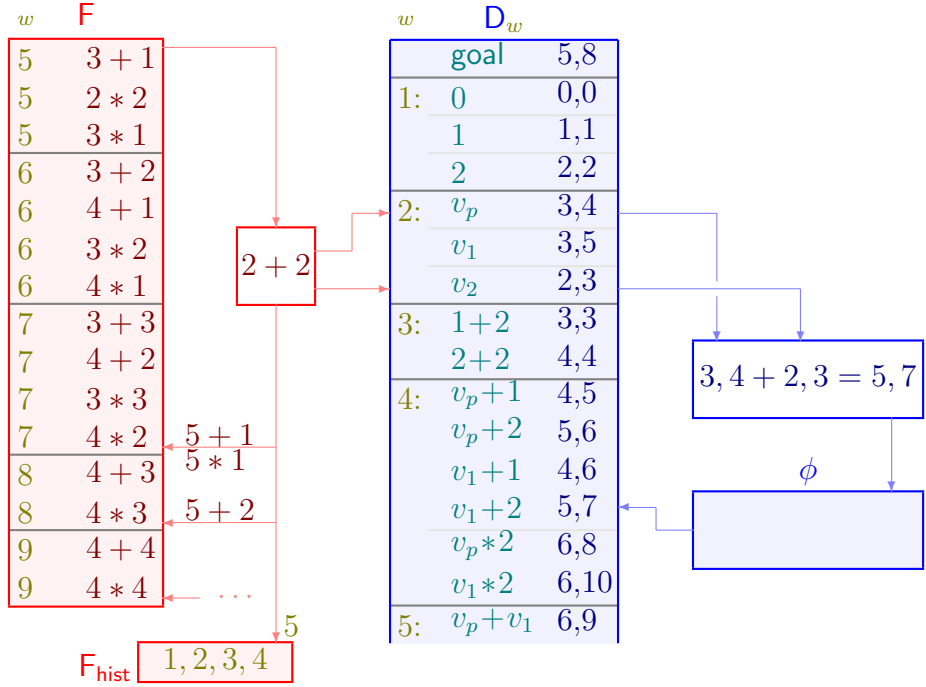
\begin{figure}

\newcommand{\0}[3]{%
	\put( 58,#1){\makebox(0,0)[bl]{\textcolor{coTerm}{$#3$}}}%
	\put( 72,#1){\makebox(0,0)[bl]{\textcolor{coVal}{#2}}}%
}

\newcommand{\1}[3]{%
	\put(  1,#1){\makebox(0,0)[bl]{\textcolor{coWg}{#2}}}%
	\put( 10,#1){\makebox(0,0)[bl]{\textcolor{coWgTerm}{$#3$}}}%
}

\newcommand{\2}[3]{%
	\put( 23,#1){\makebox(0,0)[bl]{\textcolor{coWgTerm}{$#3$}}}%
}

\begin{center}
\begin{picture}(120,90)

\thicklines

\put(  0,  10){\makebox(0,0)[bl]{\color{coHeapBg}\rule{20mm}{75mm}}}
\put(  1, 87){\makebox(0,0)[bl]{\color{coWg}$\scriptstyle w$}}
\put( 10, 87){\makebox(0,0)[b]{\color{coHeapFg}$\F$}}
      \1{ 81}{ 5}{3+1}
      \1{ 76}{ 5}{2*2}
      \1{ 71}{ 5}{3*1}
\put(  0, 70){\color{coWgSep}\line(1,0){20}}
      \1{ 66}{ 6}{3+2}
      \1{ 61}{ 6}{4+1}
      \1{ 56}{ 6}{3*2}
      \1{ 51}{ 6}{4*1}
\put(  0, 50){\color{coWgSep}\line(1,0){20}}
      \1{ 46}{ 7}{3+3}
      \1{ 41}{ 7}{4+2}
      \1{ 36}{ 7}{3*3}
      \1{ 31}{ 7}{4*2}
      \2{ 31}{ 7}{5+1}
      \2{ 28}{ 7}{5*1}
\put(  0, 30){\color{coWgSep}\line(1,0){20}}
      \1{ 26}{ 8}{4+3}
      \1{ 21}{ 8}{4*3}
      \2{ 21}{ 8}{5+2}
\put(  0, 20){\color{coWgSep}\line(1,0){20}}
      \1{ 16}{ 9}{4+4}
      \1{  11}{ 9}{4*4}
\put(  0,  10){\color{coHeapFg}\framebox(20, 75){}}

\put(17,0){\makebox(0,0)[bl]{\color{coHeapBg}\rule{20mm}{5mm}}}
\put(17,0){\color{coHeapFg}\framebox(20,5){}}
\put(17,0){\makebox(20,5){\color{coWg}$1, 2, 3, 4$}}
\put(16,0){\makebox(0,0)[br]{\color{coHeapFg}$\Fhist$}}
\put(36,6){\makebox(0,0)[bl]{\color{coWg}$5$}}

\put( 30, 50){\color{coHeapFg}\framebox(10,10){\color{coWgTerm}$2+2$}}

\put( 51, 87){\makebox(0,0)[bl]{\color{coWg}$\scriptstyle w$}}
\put( 65, 86){\makebox(0,0)[b]{\color{coDoneFg}$\D_{\textcolor{coWg}{w}}$}}
\put( 50,  5){\makebox(0,0)[bl]{\color{coDoneBg}\rule{30mm}{80mm}}}
      \0{ 81}{5,8}{\mbox{\sf goal}}
\put( 50, 80){\color{coWgSep}\line(1,0){30}}	
\put( 51, 76){\makebox(0,0)[bl]{\color{coWg}1:}}
      \0{ 76}{0,0}{0}
\put( 55, 75){\color{coOpSep}\line(1,0){25}}
      \0{ 71}{1,1}{1}
\put( 55, 70){\color{coOpSep}\line(1,0){25}}
      \0{ 66}{2,2}{2}
\put( 50, 65){\color{coWgSep}\line(1,0){30}}	
\put( 51, 61){\makebox(0,0)[bl]{\color{coWg}2:}}
      \0{ 61}{3,4}{v_p}
\put( 55, 60){\color{coOpSep}\line(1,0){25}}
      \0{ 56}{3,5}{v_1}
\put( 55, 55){\color{coOpSep}\line(1,0){25}}
      \0{ 51}{2,3}{v_2}
\put( 50, 50){\color{coWgSep}\line(1,0){30}}	
\put( 51, 46){\makebox(0,0)[bl]{\color{coWg}3:}}
      \0{ 46}{3,3}{1\.+2}
      \0{ 41}{4,4}{2\.+2}
\put( 50, 40){\color{coWgSep}\line(1,0){30}}	
\put( 51, 36){\makebox(0,0)[bl]{\color{coWg}4:}}
      \0{ 36}{4,5}{v_p\.+1}
      \0{ 31}{5,6}{v_p\.+2}
      \0{ 26}{4,6}{v_1\.+1}
      \0{ 21}{5,7}{v_1\.+2}
\put( 55, 20){\color{coOpSep}\line(1,0){25}}
      \0{ 16}{6,8}{v_p\.*2}
      \0{  11}{6,10}{v_1\.*2}
\put( 50,  10){\color{coWgSep}\line(1,0){30}}	
\put( 51,  6){\makebox(0,0)[bl]{\color{coWg}5:}}
      \0{  6}{6,9}{v_p\.+v_1}
\put( 50,  5){\color{coDoneFg}\line(0,1){80}}
\put( 50, 85){\color{coDoneFg}\line(1,0){30}}
\put( 80, 85){\color{coDoneFg}\line(0,-1){80}}

\put( 90, 35){\color{coDoneFg}\framebox(30,10){%
		\color{coVal}$3,4 + 2,3 = 5,7$}}

\put(105, 26){\makebox(0,0)[b]{\color{coDoneFg}$\phi$}}
\put( 90, 15){\makebox(0,0)[bl]{\color{coDoneBg}\rule{30mm}{10mm}}}
\put( 90, 15){\color{coDoneFg}\framebox(30,10){}}

\thinlines

\put( 20, 84){\color{coHeapLn}\line(1,0){15}}
\put( 35, 84){\color{coHeapLn}\vector(0,-1){24}}
\put( 35, 50){\color{coHeapLn}\vector(0,-1){45}}
\put( 35, 31){\color{coHeapLn}\vector(-1,0){15}}
\put( 35, 21){\color{coHeapLn}\vector(-1,0){15}}
\put( 25,  11){\color{coHeapLn}\vector(-1,0){5}}
\put( 30,  11){\color{coHeapLn}\makebox(0,0)[b]{\ldots}}
\put( 40, 57){\color{coHeapLn}\line(1,0){5}}
\put( 45, 57){\color{coHeapLn}\line(0,1){5}}
\put( 45, 62){\color{coHeapLn}\vector(1,0){5}}
\put( 40, 52){\color{coHeapLn}\vector(1,0){10}}
\put( 80, 62){\color{coDoneLn}\line(1,0){13}}
\put( 93, 62){\color{coDoneLn}\line(0,-1){9}}
\put( 93, 51){\color{coDoneLn}\vector(0,-1){6}}
\put( 80, 52){\color{coDoneLn}\line(1,0){24}}
\put(104, 52){\color{coDoneLn}\vector(0,-1){7}}
\put(116, 35){\color{coDoneLn}\vector(0,-1){10}}
\put( 90, 20){\color{coDoneLn}\line(-1,0){5}}
\put( 85, 20){\color{coDoneLn}\line(0,1){3}}
\put( 85, 23){\color{coDoneLn}\vector(-1,0){5}}

\end{picture}
\end{center}
\caption{Example state in
	Alg.~\ref{Weight decomposition list generation} (red)
	and~\ref{Term generation from decomposition lists} (blue)}
\LABEL{Example state in Alg. (red) and  (blue)}
\end{figure}

\EXAMPLE{Fibonacci sequence}{%
\LABEL{Fibonacci sequence}%
Figure~\ref{Example state in Alg. (red) and  (blue)}
shows an example state of
Alg.~\ref{Weight decomposition list generation}
and~\ref{Term generation from decomposition lists},
employed to obtain a law for the sequence $1, 2, 3; 5, 8$.
~
It assumes the $\sz$ weight from
Exm.~\ref{Weight function examples}, slightly modified by assigning
weight $2$, rather than $1$, to variable symbols.

In the left part, in red,
the Alg.~\ref{Weight decomposition list generation}
state is shown, with weight decomposition lists given in infix notation.
~
The list $\tpl{+,2,2}$ was just drawn from $\F$.
~
It evaluates to weight $5$, which occurs for the first time,
so all lists buildable from $5$ and some member of $\Fhist$ are
entered into $\F$, of which $\tpl{+,5,1}$, $\tpl{*,5,1}$, and
$\tpl{+,5,2}$ are shown as examples.

In the right part, in blue,
the Alg.~\ref{Term generation from decomposition lists}
state is shown.
~
To the right of each term $t$ in some $\D_w$,
its evaluation value under $\sigma_1$ and $\sigma_2$,
i.e.\ $\nf(t\sigma_1),\nf(t\sigma_2)$ is given;
note that both values agree for ground terms.
~
Alg.~\ref{Term generation from decomposition lists}
is just building all terms corresponding to the input weight
decomposition list $\tpl{+,2,2}$,
i.e.\ all sums of two generated minimal terms from $\D_2$.
~
After the term $v_p+v_1$ has been built and entered into $\D_5$,
the term $v_p+v_2$ is currently under consideration.
~
It evaluates to the normal form $5$ and $7$ under $\sigma_1$ and
$\sigma_2$, respectively.
~
Looking up the pair $\tpl{5,7}$ with $\phi$ reveals that there was
already a term of less weight with these values, viz.\ $v_1+2$;
so the term $v_p+v_2$ is discarded.
~
Next, the term $v_1+v_2$ will be built, evaluating to $\tpl{5,8}$;
lookup via $\phi$ will show that this is the goal pair, and the
algorithm will terminate successfully, with $v_1+v_2$ as a law term for
the given sequence.
~
We tacitly ignored commutative variants of buildable terms,
such as $v_1+v_p$, $v_2+v_p$,
and $v_2+v_1$; see Sect.~\ref{Pruning for binary operators}
below for a formal treatment.

Figure~\ref{Run of Alg. Weight decomposition list generation in Exm. Fibonacci sequence}
shows the detailed run of Alg.~\ref{Weight decomposition list generation} in this example,
up to the state shown in 
Fig.~\ref{Example state in Alg. (red) and  (blue)} (left).
~
Figures~\ref{Run of Alg. Term generation from decomposition lists in Exm. Fibonacci sequence (part 1)}
and~\ref{Run of Alg. Term generation from decomposition lists in Exm. Fibonacci sequence (part 2)}
show the full run of Alg.~\ref{Term generation from decomposition lists}, 
until the solution $v_1+v_2$ is found.
~
For sake of brevity we again ignored
commutative variants and some other trivial computations 
(marked ``skipped'' in the step field).
\qed
}

\begin{figure}
\begin{center}
{
\color{coHeapFg}
\small
\renewcommand{\arraystretch}{1.0}
\begin{tabular}{l|l|}
\cline{2-2}
3	& $\F = \set{v_p,v_1,v_2,0,1,2}$	\\
4	& $\Fhist = \set{}$	\\
\cline{2-2}
5	& $\F = \set{0,1,2,v_p,v_1,v_2}$	\\
5a	& draw $0$ from $\F$, let $w = \wg{0} = 1$	\\
5b	& \textcolor{coDoneFg}{output $0$} to Alg.12	\\
5c	& $\Fhist = \set{1}$	\\
5d	& enter $1\.+1$, $1\.*1$ into $\F$	\\
\cline{2-2}
5	& $\F = \set{1,2,v_p,v_1,v_2,1\.+1,1\.*1}$	\\
5a	& draw $1$ from $\F$, let $w = \wg{1} = 1$	\\
5b	& \textcolor{coDoneFg}{output $1$}	\\
\cline{2-2}
5	& $\F = \set{2,v_p,v_1,v_2,1\.+1,1\.*1}$	\\
5a	& draw $2$ from $\F$, let $w = \wg{2} = 1$	\\
5b	& \textcolor{coDoneFg}{output $2$}	\\
\cline{2-2}
5	& $\F = \set{v_p,v_1,v_2,1\.+1,1\.*1}$	\\
5a	& draw $v_p$ from $\F$, let $w = \wg{v_p} = 2$	\\
5b	& \textcolor{coDoneFg}{output $v_p$}	\\
5c	& $\Fhist = \set{1,2}$	\\
5d	& enter $2\.+1, 2\.*1, 2\.+2, 2\.*2$ into $\F$	\\
\cline{2-2}
5	& $\F = \set{v_1,v_2,1\.+1,1\.*1,2\.+1,2\.*1,2\.+2,2\.*2}$	\\
5a	& draw $v_1$ from $\F$, let $w = \wg{v_1} = 2$	\\
5b	& \textcolor{coDoneFg}{output $v_1$}	\\
\cline{2-2}
5	& $\F = \set{v_2,1\.+1,1\.*1,2\.+1,2\.*1,2\.+2,2\.*2}$	\\
5a	& draw $v_2$ from $\F$, let $w = \wg{v_2} = 2$	\\
5b	& \textcolor{coDoneFg}{output $v_2$}	\\
\cline{2-2}
5	& $\F = \set{1\.+1,1\.*1,2\.+1,2\.*1,2\.+2,2\.*2}$	\\
5a	& draw $1\.+1$ from $\F$, let $w = 1 \mathop{\wg{+}} 1 = 3$	\\
5b	& \textcolor{coDoneFg}{output $1\.+1$}	\\
5c	& $\Fhist = \set{1,2,3}$	\\
5d	& enter $3\.+1,3\.*1,3\.+2,3\.*2,3\.+3,3\.*3$ into $\F$	\\
\cline{2-2}
5	& $\F = \set{1\.*1,2\.+1,2\.*1,2\.+2,3\.+1,2\.*2,3\.*1,
	3\.+2,3\.*2,3\.+3,3\.*3}$	\\
5a	& draw $1\.*1$ from $\F$, let $w = 1 \mathop{\wg{*}} 1 = 3$	\\
5b	& \textcolor{coDoneFg}{output $1\.*1$}	\\
\cline{2-2}
5	& $\F = \set{2\.+1,2\.*1,2\.+2,3\.+1,2\.*2,3\.*1,
	3\.+2,3\.*2,3\.+3,3\.*3}$	\\
5a	& draw $2\.+1$ from $\F$, let $w = 2 \mathop{\wg{+}} 1 = 4$	\\
5b	& \textcolor{coDoneFg}{output $2\.+1$}	\\
5c	& $\Fhist = \set{1,2,3,4}$	\\
5d	& enter $4\.+1,4\.*1,4\.+2,4\.*2,4\.+3,4\.*3,4\.+4,4\.*4$ 
	into $\F$	\\
\cline{2-2}
5	& $\F = \set{2\.*1,2\.+2,
	\ldots, 
	3\.+2,4\.+1,3\.*2,4\.*1,
	3\.+3,4\.+2,3\.*3,4\.*2,4\.+3,4\.*3,4\.+4,4\.*4}$	\\
5a	& draw $2\.*1$ from $\F$, let $w = 2 \mathop{\wg{*}} 1 = 4$	\\
5b	& \textcolor{coDoneFg}{output $2\.*1$}	\\
\cline{2-2}
5	& $\F = \set{2\.+2, 
	\ldots,
	3\.+2,4\.+1,3\.*2,4\.*1,
	3\.+3,4\.+2,3\.*3,4\.*2,4\.+3,4\.*3,4\.+4,4\.*4}$	\\
5a	& draw $2\.+2$ from $\F$, let $w = 2 \mathop{\wg{+}} 2 = 5$	\\
5b	& \textcolor{coDoneFg}{output $2\.+2$}	\\
	& \textcolor{black}{------state shown in 
	Fig.~\ref{Example state in Alg. (red) and (blue)} (left)------}	\\
5c	& $\Fhist = \set{1,2,3,4,5}$	\\
5d	& enter 
	$5\.+1,5\.*1,5\.+2,5\.*2,5\.+3,5\.*3,5\.+4,5\.*4,5\.+5,5\.*5$ 
	into $\F$	\\
\cline{2-2}
\end{tabular}
}
\caption{Run of Alg.~\ref{Weight decomposition list generation} 
	in Exm.~\ref{Fibonacci sequence}}
\LABEL{Run of Alg. Weight decomposition list generation in Exm. Fibonacci sequence}
\end{center}
\end{figure}

\begin{figure}
\begin{center}
{
\color{coDoneFg}
\small
\renewcommand{\arraystretch}{1.0}
\begin{tabular}{l|l|}
\cline{2-2}
3	& $\phi = \set{}$	\\
4	& $\D_1 = \D_2 = \ldots = \set{}$	\\
\cline{2-2}
5a	& \textcolor{coHeapFg}{read $0$} from Alg.11, 
	let $w = \wg{0} = 1$	\\
5bi,ii	& build the term $0$, it evaluates to $0$, $0$	\\
5biiiA,B	& add $\tpl{0,0} \mapsto 0$ to $\phi$, 
	add $0$ to $\D_1$	\\
\cline{2-2}
5a	& \textcolor{coHeapFg}{read $1$}, let $w = \wg{1} = 1$	\\
5bi,ii	& build the term $1$, it evaluates to $1$, $1$	\\
5biiiA,B	& add $\tpl{1,1} \mapsto 1$ to $\phi$, 
	add $1$ to $\D_1$	\\
\cline{2-2}
5a	& \textcolor{coHeapFg}{read $2$}, let $w = \wg{2} = 1$	\\
5bi,ii	& build the term $2$, it evaluates to $2$, $2$	\\
5biiiA,B	& add $\tpl{2,2} \mapsto 2$ to $\phi$, 
	add $2$ to $\D_1$	\\
\cline{2-2}
5a	& \textcolor{coHeapFg}{read $v_p$}, let $w = \wg{v_p} = 2$	\\
5bi,ii	& build the term $v_p$, it evaluates to $3$, $4$	\\
5biiiA,B	& add $\tpl{3,4} \mapsto v_p$ to $\phi$, 
	add $v_p$ to $\D_2$	\\
\cline{2-2}
5a	& \textcolor{coHeapFg}{read $v_1$}, let $w = \wg{v_1} = 2$	\\
5bi,ii	& build the term $v_1$, it evaluates to $3$, $5$	\\
5biiiA,B	& add $\tpl{3,5} \mapsto v_1$ to $\phi$, 
	add $v_1$ to $\D_2$	\\
\cline{2-2}
5a	& \textcolor{coHeapFg}{read $v_2$}, let $w = \wg{v_2} = 2$	\\
5bi,ii	& build the term $v_2$, it evaluates to $2$, $3$	\\
5biiiA,B	& add $\tpl{2,3} \mapsto v_2$ to $\phi$, 
	add $v_2$ to $\D_2$	\\
\cline{2-2}
5a	& \textcolor{coHeapFg}{read $1\.+1$}, 
	let $w = 1 \mathop{\wg{+}} 1  = 3$	\\
5b	& combine with ``$+$'' all terms 
	from $\D_1 = \set{0,1,2}$ and $\D_1$:	\\
5bi,ii	& build the term $0\.+0$, it evaluates to $0$, $0$	\\
5biii	& $\phi(0,0) = 0$ is already defined	\\
5bi,ii	& build the term $0\.+1$, it evaluates to $1$, $1$	\\
5biii	& $\phi(1,1) = 1$ is already defined	\\
5bi,ii	& build the term $0\.+2$, it evaluates to $2$, $2$	\\
5biii	& $\phi(2,2) = 2$ is already defined	\\
5bi,ii	& build the term $1\.+0$, it evaluates to $1$, $1$	\\
5biii	& $\phi(1,1) = 1$ is already defined	\\
5bi,ii	& build the term $1\.+1$, it evaluates to $2$, $2$	\\
5biii	& $\phi(2,2) = 2$ is already defined	\\
5bi,ii	& build the term $1\.+2$, it evaluates to $3$, $3$	\\
5biiiA,B	& add $\tpl{3,3} \mapsto 1\.+2$ to $\phi$, 
	add $1\.+2$ to $\D_3$	\\
5bi,ii	& build the term $2\.+0$, it evaluates to $2$, $2$	\\
5biii	& $\phi(2,2) = 2$ is already defined	\\
5bi,ii	& build the term $2\.+1$, it evaluates to $3$, $3$	\\
5biii	& $\phi(3,3) = 1\.+2$ is already defined	\\
5bi,ii	& build the term $2\.+2$, it evaluates to $4$, $4$	\\
5biiiA,B	& add $\tpl{4,4} \mapsto 2\.+2$ to $\phi$, 
	add $2\.+2$ to $\D_3$	\\
\cline{2-2}
5a	& \textcolor{coHeapFg}{read $1\.*1$}, 
	let $w = 1 \mathop{\wg{*}} 1 = 3$	\\
5bi	& build the terms 
	$0\.*0,0\.*1,0\.*2, 1\.*0,1\.*1,1\.*2, 2\.*0,2\.*1,2\.*2$	\\
skipped	& none of them evaluate to a new value vector, 
	so no change results	\\
\cline{2-2}
\end{tabular}
}
\caption{Run of Alg.~\ref{Term generation from decomposition lists} 
	in Exm.~\ref{Fibonacci sequence} (part 1)}
\LABEL{Run of Alg. Term generation from decomposition lists in Exm. Fibonacci sequence (part 1)}
\end{center}
\end{figure}

\begin{figure}
\begin{center}
{
\color{coDoneFg}
\small
\renewcommand{\arraystretch}{1.0}
\begin{tabular}{l|l|}
\cline{2-2}
5a	& \textcolor{coHeapFg}{read $2\.+1$}, 
	let $w = 2 \mathop{\wg{+}} 1 = 4$	\\
5b	& combine with ``$+$'' all terms 
	from $\D_2 = \set{v_p,v_1,v_2}$ and $\D_1 = \set{0,1,2}$:	\\
skipped	& terms $v_p\.+0,v_1\.+0,v_2\.+0$ cause no change	\\
5bi,ii	& build the term $v_p\.+1$, it evaluates to $4$, $5$	\\
5biiiA,B	& add $\tpl{4,5} \mapsto v_p\.+1$ to $\phi$, 
	add $v_p\.+1$ to $\D_4$	\\
5bi,ii	& build the term $v_p\.+2$, it evaluates to $5$, $6$	\\
5biiiA,B	& add $\tpl{5,6} \mapsto v_p\.+2$ to $\phi$, 
	add $v_p\.+2$ to $\D_4$	\\
5bi,ii	& build the term $v_1\.+1$, it evaluates to $4$, $6$	\\
5biiiA,B	& add $\tpl{4,6} \mapsto v_1\.+1$ to $\phi$, 
	add $v_1\.+1$ to $\D_4$	\\
5bi,ii	& build the term $v_1\.+2$, it evaluates to $5$, $7$	\\
5biiiA,B	& add $\tpl{5,7} \mapsto v_1\.+2$ to $\phi$, 
	add $v_1\.+2$ to $\D_4$	\\
5bi,ii	& build the term $v_2\.+1$, it evaluates to $3$, $4$	\\
5biii	& $\phi(3,4) = v_p$ is already defined	\\
5bi,ii	& build the term $v_2\.+2$, it evaluates to $4$, $5$	\\
5biii	& $\phi(4,5) = v_p\.+1$ is already defined	\\
\cline{2-2}
5a	& \textcolor{coHeapFg}{read $2\.*1$}, 
	let $w = 2 \mathop{\wg{*}} 1 = 4$	\\
5b	& combine with ``$*$'' all terms from 
	$\D_2 = \set{v_p,v_1,v_2}$ and $\D_1 = \set{0,1,2}$:	\\
skipped	& terms $v_p\.*0,v_p\.*1,v_1\.*0,v_1\.*1,v_2\.*0,v_2\.*1$ 
	cause no change	\\
5bi,ii	& build the term $v_p\.*2$, it evaluates to $6,8$	\\
5biiiA,B	& add $\tpl{6,8} \mapsto v_p\.*2$ to $\phi$, 
	add $v_p\.*2$ to $\D_4$	\\
5bi,ii	& build the term $v_1\.*2$, it evaluates to $6,10$	\\
5biiiA,B	& add $\tpl{6,10} \mapsto v_1\.*2$ to $\phi$, 
	add $v_1\.*2$ to $\D_4$	\\
5bi,ii	& build the term $v_2\.*2$, it evaluates to $4,6$	\\
5biii	& $\phi(4,6) = v_1\.+1$	\\
\cline{2-2}
5a	& \textcolor{coHeapFg}{read $2\.+2$}, 
	let $w = 2 \mathop{\wg{+}} 2 = 5$	\\
5b	& combine with ``$+$'' all terms from 
	$\D_2 = \set{v_p,v_1,v_2}$ and $\D_2$:	\\
skipped	& terms $v_p\.+v_p,v_1\.+v_1,v_2\.+v_2$ cause no change	\\
5bi,ii	& build the term $v_p\.+v_1$, it evaluates to $6$, $9$	\\
5biiiA,B	& add $\tpl{6,9} \mapsto v_p\.+v_1$ to $\phi$, 
	add $v_p\.+v_1$ to $\D_5$	\\
5bi,ii	& build the term $v_p\.+v_2$, it evaluates to $5$, $7$	\\
5biii	& $\phi(5,7) = v_1\.+2$ is already defined	\\
	& \textcolor{black}{------state shown in
	Fig.~\ref{Example state in Alg. (red) and (blue)} (right)------} \\
5bi,ii	& build the term $v_1\.+v_2$, it evaluates to $5$, $8$	\\
5biiiA,B	& add $\tpl{5,8} \mapsto v_1\.+v_2$ to $\phi$, 
	add $v_1\.+v_2$ to $\D_5$	\\
5biiiC	& found $\tpl{5,8}$, 
	so \textcolor{black}{\bf stop} with success	\\
\cline{2-2}
\end{tabular}
}
\caption{Run of Alg.~\ref{Term generation from decomposition lists} 
	in Exm.~\ref{Fibonacci sequence} (part 2)}
\LABEL{Run of Alg. Term generation from decomposition lists in Exm. Fibonacci sequence (part 2)}
\end{center}
\end{figure}

\cleardoublepage

\subsection{Correctness and completeness}
\LABEL{Correctness and completeness}

\LEMMA{Properties of Alg.~\ref{Weight decomposition list generation}}{%
\LABEL{Properties of Alg. Weight decomposition list generation}%
Algorithm~\ref{Weight decomposition list generation} has the following
properties.
\begin{enumerate}
\item
	\LABEL{Properties of Alg. Weight decomposition list generation
	/ no duplicates}
	No Weight decomposition list appears twice in the output of
	Alg.~\ref{Weight decomposition list generation}.
\item
	\LABEL{Properties of Alg. Weight decomposition list generation
	/ nondecreasing}
	The sequence of evaluating weights of output decomposition lists
	is non-decreasing.
\item
	\LABEL{Properties of Alg. Weight decomposition list generation 
	/ nonstationary}
	This sequence cannot get stationary infinitely long.
\item
	\LABEL{Properties of Alg. Weight decomposition list generation 
	/ Liveness}
	Each weight decomposition list evaluating to a weight $< \infty$
	that is inserted into $\F$ (in step~\ref{Weight-gen heap init}
	or~\ref{Weight-gen heap insert})
	will eventually be drawn out of it (in step~\ref{Weight-gen draw}).
\end{enumerate}
}
\PROOF{
\begin{enumerate}
\item
	Each weight decomposition list entered into $\F$
	in step~\ref{Weight-gen heap insert} contains a new weight,
	viz.\ $w$ from step~\ref{Weight-gen draw},
	among its arguments.
\item
	In step~\ref{Weight-gen heap insert}
	each weight decomposition list
	inserted into $\F$
	evaluates to a weight greater or equal to the current
	minimum, due to the monotonicity of weight functions.
\item
	It suffices to
	show that there are only finitely many decomposition
	lists evaluating to the same weight.
	~
	Given  some $\tpl{f,\:,1n{x_\i}}$, 
	let $x = \wg{f}(\:,1n{x_\i})$ be the weight it evaluates to.
	~
	Since $(<)$ is well-founded,
	there are only finitely many weights $\leq x$. 
	~
	However, each decomposition list $\tpl{g,\:,1n{y_\i}}$
	with $y_i > x$ for some $i$ evaluates to a weight $> x$
	since $\wg{g}$ is increasing.
	~
	(Totality of $(<)$ is needed for this argument, but no strict
	increasingness of $\wg{g}$.)
\item
	Assume the weight decomposition list $\tpl{f,\:,1n{w_\i}}$
	has been inserted into $\F$.
	~
	We show that it will eventually be drawn from $\F$.
	~
	Consider again the the sequence of evaluating weights of
	output decomposition lists.
	~
	If it is a finite sequence, $\F$ must eventually get empty,
	and hence
	$\tpl{f,\:,1n{w_\i}}$ must have been drawn from $\F$.
	~
	If it is an infinite one, it will eventually get larger than the
	result weight of $\tpl{f,\:,1n{w_\i}}$,
	by~\ref{Properties of Alg. Weight decomposition list generation
	/ nondecreasing}
	and~\ref{Properties of Alg. Weight decomposition list generation
	/ nonstationary};
	before this can happen, $\tpl{f,\:,1n{w_\i}}$ must have been drawn
	from $\F$,
	by~\ref{Properties of Alg. Weight decomposition list generation
	/ nondecreasing}.
\qed
\end{enumerate}
}

\LEMMA{Completeness of Alg.~\ref{Weight decomposition list generation}}{%
\LABEL{Completeness of weight decomposition list generation}%
Let $V$ be the set of variables given to
Alg.~\ref{Weight decomposition list generation}.
~
Let $t \in \T{}{}{V}$ be an arbitrary term in these variables,
let $t = f( \:,1n{t_\i} )$.
~
Then the output stream of
Alg.~\ref{Weight decomposition list generation}
will eventually contain the list
$\tpl{f, \:,1n{\WG(t_\i)}}$; this list evaluates to weight $\WG(t)$.
~
Observe that for $n=0$ also constants and variables are admitted as $t$.
}
\PROOF{
By 
Lem.~\ref{Properties of Alg. Weight decomposition list generation}.%
\ref{Properties of Alg. Weight decomposition list generation / Liveness}
it is sufficient to show that
$\tpl{f, \:,1n{\WG(t_\i)}}$ is eventually entered into $\F$.
~
We do this by induction on $t$:
\begin{itemize}
\item If $n=0$, i.e.\ if $t$ is a constant or variable,
	it is entered into $\F$ in step~\ref{Weight-gen heap init}
	of Alg.~\ref{Weight decomposition list generation}.
\item If $n>0$, then by induction hypothesis (I.H.)
	some list evaluating to $\WG(t_i)$
	will eventually be drawn from $\F$, for $i=\:,1n\i$.
	~
	When the last, i.e.\ the largest of them is drawn for the
	first time in step~\ref{Weight-gen draw},
	we have $\set{ \:,1n{\WG(t_\i)}} \subseteq \Fhist$.
	~
	Therefore,
	the weight decomposition list $\tpl{f, \:,1n{\WG(t_\i)}}$
	is among those that are entered into $\F$ in
	step~\ref{Weight-gen heap insert}.
\qed
\end{itemize}
}

\LEMMA{Value-pair cache property}{%
\LABEL{Value-pair cache property}%
Let $t$ be the term built in step~\ref{cached term-gen build} of
Alg.~\ref{Term generation from decomposition lists}.
~
After completion of step~\ref{cached term-gen new},
$\phi(\nf(t \sigma_1),\nf(t \sigma_2))$ is defined, and yields
a term $t'$ such that
$t' \sigma_1 =_E t \sigma_1 \land t' \sigma_2 =_E t \sigma_2$
and $\WG(t') \leq \WG (t)$.
}
\PROOF{
In step~\ref{cached term-gen eval},
Alg.~\ref{Term generation from decomposition lists}
computes $v_1 = \nf(t \sigma_1)$ and $v_2 = \nf(t \sigma_2)$.
~
When $\phi(v_1,v_2)$ is yet undefined, it is set to $t$ and there is
nothing to show.

When $\phi(v_1,v_2)$ was already defined,
say to be $t'$,
we have $\WG(t') \leq \WG (t)$,
since the terms in step~\ref{cached term-gen build} are built in
order of non-decreasing weight
(a property of Alg.~\ref{Weight decomposition list generation}).
~
Moreover, we have
$t' \sigma_1
= \phi(v_1,v_2) \sigma_1
=_E v_1
=_E t \sigma_1$
using Def.~\ref{Equality, normal form},
and similar for $\sigma_2$.
\qed
}

\LEMMA{}{%
\LABEL{Completeness of value-pair cached term generation 1}%
In the setting of Alg.~\ref{Value-pair cached term generation},
let $V = \dom \sigma_1 \cup \dom \sigma_2$.
~
If  the test in step~\ref{cached term-gen goal} of
Alg.~\ref{Term generation from decomposition lists} 
is omitted such that the
loop of step~\ref{cached term-gen empty} processes all weight
decomposition lists, then the thus modified 
Alg.~\ref{Value-pair cached term generation} has the following property:
~
For each term $t \in \T{}{}{V}$,
a term $t'$
with $t' \sigma_1 =_E t \sigma_1$ and $t' \sigma_2 =_E t \sigma_2$
is eventually entered into $\D_{\WG(t')}$ such that $\WG(t') \leq \WG(t)$.
}
\PROOF{
Induction on $t$;
let $t = f( \:,1n{t_\i} )$, where $n$ may also be zero.
~
By I.H., the algorithm will eventually find terms $t'_i$ such that
$t'_i \sigma_1 =_E t_i \sigma_1 \land t'_i \sigma_2 =_E t_i \sigma_2$
and $\WG(t'_i) \leq \WG(t_i)$,
and put them into $\D_{\WG(t'_i)}$, for $i=\:,1n\i$.

By Lem.~\ref{Completeness of weight decomposition list generation},
eventually a weight decomposition list
$\tpl{f, \:,1n{\WG(t'_\i)} }$ will appear in the stream in
step~\ref{cached term-gen draw}
of Alg.~\ref{Term generation from decomposition lists}.
~
The algorithm will then generate $f(\:,1n{t''_\i})$
for all $t''_i \in \D_{\WG(t'_i)}$, for $i=\:,1n\i$.
~
In particular, it will build
the term $f(\:,1n{t'_\i})$ in step~\ref{cached term-gen build}.

By Lem.~\ref{Value-pair cache property},
after completion of algorithm step~\ref{cached term-gen new}
the entry
$\phi(f(\:,1n{t'_\i}) \sigma_1,f(\:,1n{t'_\i}) \sigma_2)$
is defined, say to be $t'$,
with the property
$t' \sigma_1
=_E f(\:,1n{t'_\i}) \sigma_1
=_E f(\:,1n{t_\i}) \sigma_1
= t \sigma_1$.
~
Similarly, we have $t' \sigma_2 =_E t \sigma_2$.
~
Moreover, Lem.~\ref{Value-pair cache property} yields that
$\WG(t')
\leq \WG(f(\:,1n{t'_\i}))
\leq \WG(f(\:,1n{t_\i}))
= \WG(t)$, using $\wg{f}$'s monotonicity.
\qed
}

\THEOREM{Correctness and completeness 
	of Alg.~\ref{Value-pair cached term generation}}{%
\LABEL{Correctness and completeness of value-pair cached term generation}%
Let $g_1,g_2 \in \NF$ be the goal values given to
Alg.~\ref{Term generation from decomposition lists}.
~
If some term $t$ exists such that
\\
\rule{0cm}{0cm}
\hfill $t \sigma_1 =_E g_1 \land t \sigma_2 =_E g_2$,
\hfill $(*)$
\\
then Alg.~\ref{Term generation from decomposition lists}
will eventually stop successfully in step~\ref{cached term-gen goal},
with a result term that is of minimal weight with property $(*)$.
~
Note that the result term needn't be $t$ itself.
~
If no such term $t$ exists, the algorithm will fail in
step~\ref{cached term-gen fail}, or will loop forever.
}
\PROOF{
W.l.o.g., let $t$ be a term of minimal weight with property $(*)$.
~
By Lem~\ref{Completeness of value-pair cached term generation 1},
we have that the modified
(test step~\ref{cached term-gen goal} omitted)
Alg.~\ref{Term generation from decomposition lists}
will eventually generate a term $t'$ with property $(*)$ such that
$\WG(t') \leq \WG(t)$.
~
Since $t$ was minimal, we get $\WG(t') = \WG(t)$ also minimal.

Hence, in the unmodified algorithm,
test step~\ref{cached term-gen goal} will apply to $t'$, unless it
applied to an earlier-generated term.
~
In both cases, we are done, since the test ensures property $(*)$,
and the first term passing the test is of minimal weight
(again, a property of
Alg.~\ref{Weight decomposition list generation}).
~
Note that
Lem.~\ref{Completeness of value-pair cached term generation 1}
and the existence of $t$
ensures that Alg.~\ref{Term generation from decomposition lists}
doesn't stop with failure in step~\ref{cached term-gen fail}.

If no term $t$ exists with property $(*)$, the test in
step~\ref{cached term-gen goal} cannot succeed.
~
Hence, the only way to terminate the algorithm is in
step~\ref{cached term-gen fail}.
\qed
}

\EXAMPLE{Incompleteness for proper weight limit ordinal}{%
If $\W$ and $(<)$ were such that a limit ordinal different from
$\infty$ existed in $\W$, then
Alg.~\ref{Value-pair cached term generation}
may be incomplete.
~
Modifying Def.~\ref{Weight domain, minimal weight},
let $\W = (\N \times \N) \cup \set{ \infty }$,
ordered by the lexicographic combination of the usual order on natural
numbers, with each proper pair being less than $\infty$.
~
Consider the signature $\Sigma = \set{ 3, (+), (*) }$.
~
Define weight functions by
\[\begin{array}{rcl c rl l}
& \wg{3} && = & \tpl{0, & 1}, \\
\tpl{w_1,w_2} & \wg{+} & \tpl{w_3,w_4}
	& = & \tpl{1+w_1+w_3, & 1+w_2+w_4}, \\
\tpl{w_1,w_2} & \wg{*} & \tpl{w_3,w_4}
	& = & \tpl{w_1+w_3, & 1+w_2+w_4},
	& \mbox{ and} \\
\multicolumn{5}{l}{\infty \wg{+} w
	= w \wg{+} \infty
	= \infty \wg{*} w
	= w \wg{*} \infty
	= \infty.}	\\
\end{array}\]
Intuitively, we have $\WG(t)$ being
a pair consisting of the number of $(+)$-occurrences in
$t$ and the total number of symbols in $t$.
~
All these weight functions are strictly monotonic, strictly
increasing, and strict.
\begin{itemize}
\item $\wg{*}$ is strictly monotonic:
	\begin{itemize}
	\item Let $\tpl{w_1,w_2} < \tpl{w'_1,w'_2}$.
		\begin{itemize}
		\item If $w_1 < w'_1$,
			then $w_1+w_3 < w'_1+w_3$.
		\item If $w_1 = w'_1$ and $w_2 < w'_2$,
			then $1+w_2+w_4 < 1+w'_2+w_4$.
		\end{itemize}
		From any case, we get
		$\tpl{w_1,w_2} \wg{*} \tpl{w_3,w_4}
		= \tpl{w_1+w_3,1+w_2+w_4}
		< \tpl{w'_1+w_3,1+w'_2+w_4}
		= \tpl{w'_1,w'_2} \wg{*} \tpl{w_3,w_4}$.
	\item If $w_{12} < \infty$,
		then
		$w_{12} \wg{*} w_{34}
		\leq \infty
		= \infty \wg{*} w_{34}$.
	\end{itemize}
\item $\wg{*}$ is strictly increasing:
	\begin{itemize}
	\item \begin{itemize}
		\item If $w_3 = 0$,
			then $w_2 < 1+w_2+w_4$.
		\item If $w_3 > 0$,
			then $w_1 < w_1+w_3$.
		\end{itemize}
		From any case, we get	\\
		$\tpl{w_1,w_2}
		< \tpl{w_1+w_3, 1+w_2+w_4}
		= \tpl{w_1,w_2} \wg{*} \tpl{w_3,w_4}$.
	\item $\infty \leq \infty = \infty \wg{*} w_{34}$
		is trivial.
	\end{itemize}
\item $\wg{+}$ similar.
\end{itemize}

There is an infinite ascending chain of weights
$\:<1{2n\.+1}{\tpl{0,\i}} < \ldots < \tpl{1,3}$,
corresponding to a sequence of terms
$3, \; 3\.*3, \; (3\.*3)\.*3, \; \ldots, \; 3\.+3$.
~
We don't need variables and substitutions in this example;
if desired, one may think
both $\sigma_1$ and $\sigma_2$ being the empty substitution.
~
There is a term evaluating to $6$, viz.~$t = 3\.+3$,
however all infinitely many
$*$-terms of the chain are of less weight and will be
generated before $t$.
~
Since each of them denotes a power of $3$,
none may evaluate to $6$.
\qed
}

\newcommand{\R}{\mathcal{R}}
\newcommand{\NT}{\mathcal{N}}

\subsection{Relation to the grammar-based algorithm}
\LABEL{Relation to the grammar-based algorithm}

In this section, we give an argument to show that
Alg.~\ref{Value-pair cached term generation}
is more efficient than the
\notion{Constrained $E$-generalization algorithm}
from \cite[Sect.3.1, p.6]{Burghardt.2005c}.
~
The latter algorithm used regular tree grammars to represent
equivalence classes (mod.\ $E$) of values as well as sets of
generalization terms.
~
We will assume here that a more concise
representation is possible, based on \notion{grammar schemes}
(as suggested in~\cite[p.11]{Burghardt.2005c}).

We will make certain unjustified, but optimistic,
assumptions about grammar schemes,
their formal definability, and their implementability.
~
We will show that even under these assumptions an appropriately improved
Constrained $E$-generalization algorithm is not
better than Alg.~\ref{Value-pair cached term generation},
at least for equational theories of
realistic complexity.

A (nondeterministic) regular tree grammar
\cite{Thatcher.Wright.1968,Comon.Dauchet.Gilleron.2001}
is a triple $\G = \tpl{\Sigma,\NT,\R}$.
~
$\Sigma$ is a signature,
$\NT$ is a finite set of nonterminal symbols
and
$\R$ is a finite set of rules of the form
$N \sortdef \:{\mid}1m{ f_\i( \:,{1}{n_\i}{ N_{\i\j} } ) }$
or, abbreviated,
$N \sortdef \bigmid_{i=1}^m \; f_i(N_{i1},...,N_{in_i})$.
~
Given a grammar $\G$ and a nonterminal $N \in \NT$,
the language $\L_\G(N)$ produced by $N$
is defined in the usual way
as the set of all
ground terms derivable from $N$ as the start symbol.
~
For our purposes,
it is sufficient to define a \notion{grammar scheme}
somewhat informal to be
an algorithm that generates a grammar from some input.

We will present our argument along the following $E$-generalization
example.
~
Consider the equational theory $E$, defining $(+)$, from
Fig.~\ref{Background theory (l), sequence example (m), sequence law synthesis (r)}
(left, topmost two equations).
~
Its equivalence classes can be described by a grammar scheme $\G$
consisting of the rules
\[\begin{array}{cc c ccc @{\hspace*{1cm}} l}
N_0 & \sortdef & 0
	& \mid && N_0+N_0	\\
N_n & \sortdef & s(N_{n-1})
	& \mid & \bigmid_{i=0}^i & N_i+N_{n-i}
	& \mbox{ for } n \geq 1	\\
\end{array}\]
Due to the aggregation of the rules for $n \geq 1$,
this is a grammar scheme rather than just a grammar.
~
We have that
$\L_\G(N_n) = \eqc{s^n(0)}{}{}$  is the equivalence class of all terms that
evaluate to $s^n(0)$ wrt.~$=_E$.
~
Abbreviating again $s^n(0)$ by $n$ for convenience,
consider the example substitutions
\[\begin{array}{lllllll}
\sigma_1 & = & \subst{& v_1 \mapsto 3, & v_2 \mapsto 2& }
	& \mbox{ and}	\\
\sigma_2 & = & \subst{& v_1 \mapsto 5, & v_2 \mapsto 3& }
	& \mbox{ .}	\\
\end{array}\]
If in $\G$ we add
\begin{itemize}
\item to the rule of $N_3$ an alternative ``$\ldots \mid v_1$'',
\item to the rule of $N_2$ an alternative ``$\ldots \mid v_2$'', and
\end{itemize}
we get a grammar scheme $\G_1$
such that
$\L_{\G_1}(N_n)$
is the set of all terms $t$ such that $t \sigma_1$
evaluates to $n$,
by \cite[Theorem 7 in Sect.~1.4]{Comon.Dauchet.Gilleron.1999}.
~
Similarly,
if in $\G$ we add
\begin{itemize}
\item to the rule of $N_5$ an alternative ``$\ldots \mid v_1$'',
\item to the rule of $N_3$ an alternative ``$\ldots \mid v_2$'', and
\end{itemize}
we get a grammar scheme $\G_2$
such that $\L_{\G_2}(N_n)$ is the set of terms evaluating to
$n$ under $\sigma_2$.

Next, we will compute a grammar scheme $\G_{1,2}$ containing,
for each $n,n' \in \N$,
a nonterminal $N_{n,n'}$
with $\L_{\G_{1,2}}(N_{n,n'}) = \L_{G_1}(N_n) \cap \L_{G_2}(N_{n'})$
being the set of all terms evaluating to $n$ and $n'$ under $\sigma_1$ and
$\sigma_2$, respectively.
~
For example, we will get
$v_1+v_2 \in \L_{\G_{1,2}}(N_{5,8})$,
indicating that the term $v_1+v_2$ is a common generalization of
$5$ and $8$ w.r.t.~$(=_E)$ and the given substitutions.
~
This will allow us later on
to establish that term as a construction law of the
Fibonacci sequence, as explained in \cite[Sect.~5.2]{Burghardt.2005c}.

Following the usual product-automaton construction
(e.g.~\cite[Sect.~1.3]{Comon.Dauchet.Gilleron.1999}),
we obtain a grammar scheme $\G_{1,2}$:
\[\begin{array}{cc@{}c@{}cl clll@{}c@{}l @{\hspace*{5mm}} l}
N_{0,0}
	& \sortdef
	& 0
	&&& \mid
	&&& N_{0,0} & + & N_{0,0}	\\
N_{0,n'}
	& \sortdef
	&
	&&&&& \bigmid_{i'=0}^{n'}
	& N_{0,i'} & + & N_{0,n'-i'}
	& \mbox{ for } n' \geq 1	\\
N_{n,0}
	& \sortdef
	&
	&&&& \bigmid_{i=0}^n
	&& N_{i,0} & + & N_{n-i,0}
	& \mbox{ for } n \geq 1	\\
N_{2,3}
	& \sortdef
	& s(N_{1,2})
	& \mid
	& v_2
	& \mid
	& \bigmid_{i=0}^2 \
	& \bigmid_{i'=0}^3 \;
	& N_{i,i'} & + & N_{2\.-i,3\.-i'}	\\
N_{3,5}
	& \sortdef
	& s(N_{2,4})
	& \mid
	& v_1
	& \mid
	& \bigmid_{i=0}^3
	& \bigmid_{i'=0}^5 \;
	& N_{i,i'} & + & N_{3\.-i,5\.-i'}	\\
N_{n,n'}
	& \sortdef
	& s(N_{n\.-1,n'\.-1})
	& \mid
	&&& \bigmid_{i=0}^n
	& \bigmid_{i'=0}^{n'}
	& N_{i,i'} & + & N_{n\.-i,n'\.-i'}
	& \mbox{ for all other } n,n'	\\
\end{array}\]
Since $N_{5,8} \sortdef \ldots N_{3,5}+N_{2,3} \ldots$,
we will get in fact $v_1+v_2 \in \L_{\G_{1,2}}(N_{5,8})$.


We use Knuth's algorithm~\cite{Knuth.1977} to compute a term $t_{n,n'}
\in \L_{\G_{1,2}}(N_{n,n'})$ of minimal weight.
~
When applied to a grammar scheme like that of $\G_{1,2}$, 
and when terms are maintained along with their weights,
it amounts to the following method:

\ALGORITHM{Knuth's algorithm for grammar schemes}{%
\LABEL{Knuth's algorithm for grammar schemes}%
Input:
\begin{itemize}
\item a regular tree grammar scheme $\G$,
	with the signature $\Sigma$ being its set of terminal symbols,
	and its nonterminals having the form $N_{n,n'}$
	with $n, n' \in \N$
\item a computable weight function $\wg{f}$,
	for each function symbol $f \in \Sigma$
\item a pair $\tpl{g_1,g_2} \in \N \times \N$ of goal values
\end{itemize}

Output:
\begin{itemize}
\item a partial mapping $\phi: \NT \hookrightarrow \T{}{}{\set{}}$
	from nonterminals
	to weight-minimal terms of their generated language,
	such that $\phi(N_{n,n'})$, if defined,
	is a weight-minimal term in $\L_\G(N_{n,n'})$,
	and such that $\phi(N_{g_1,g_2})$ is defined
	if $\L_\G(N_{g_1,g_2})$ isn't empty
\end{itemize}

Algorithm:
\begin{enumerate}
\item
	Initially, let $\phi$ be the empty mapping.
\item\LABEL{knuth loop start}
	While $N_{g_1,g_2} \not\in \dom \phi$, perform
	steps~\ref{knuth apply} to~\ref{knuth add}:
	\begin{enumerate}
	\item\LABEL{knuth apply}
		For each grammar rule alternative
		$N_{n,n'} \sortdef \ldots \mid f(\:,1k{N_{i_\i,i'_\i}})$,
		such that $N_{n,n'}$ is not in $\dom \phi$,
		but all $\:,1k{N_{i_\i,i'_\i}}$ are,
		consider the term $f( \:,1k{\phi(N_{i_\i,i'_\i})})$.
	\item
		Among these considered terms,
		choose some $t$ of minimal weight.
	\item
		Let $N_{n,n'} \sortdef \ldots$ be the grammar rule
		in step~\ref{knuth apply}
		where $t$ originated from.
	\item\LABEL{knuth add}
		Add $\tpl{N_{n,n'},t}$ to $\phi$.
	\qed
	\end{enumerate}
\end{enumerate}
}

We make the optimistic assumptions that
Alg.~\ref{Knuth's algorithm for grammar schemes}
can be implemented to handle even
\begin{itemize}
\item grammar schemes representing rules for infinitely many
	nonterminals
	(such as $N_{n,n'}$ for all $n,n' \in \N$ in our current
	example), and
\item grammar scheme rules with infinitely many alternatives
	(such as
	$N_{0,0} \sortdef
	\bigmid_{i=0}^\infty
	\; \bigmid_{i'=0}^\infty
	\; N_{i,i'} - N_{i,i'}$
	for an equational theory that includes subtraction).
\end{itemize}
However, we require that there are only finitely many
function symbols, and each of them has a fixed finite arity.

Next, if sufficiently many or sufficiently sophisticated
operators are handled by the grammars,
then for every combination $j,j'$ a corresponding nonterminal $N_{j,j'}$
will appear in $\G_{1,2}$, as Exm.~\ref{Fully connected grammar}
demonstrates.

\EXAMPLE{Fully connected grammar}{%
\LABEL{Fully connected grammar}%
Let $\G$ be a tree grammar such that $\L_G(N_i) = \eqc{s^i(0)}{E}{}$
wrt.~some equational
theory describing at least $(+)$ and $(-)$; assume $\:,0n{N_\i}$ belong
to $\G$.
~
Then for each $0 \leq i , j \leq n$,
we have a derivation
\[N_i \;\raa\; N_n - N_{n-i} \;\raa\; (N_j + N_{n-j}) - N_{n-i} .\]
Hence, such derivations also exist in the lifted grammars
$\G_1$ and $\G_2$.
~
In the intersection grammar $G_{1,2}$, we therefore have a derivation
\[N_{i,i'} \;\raa\; N_{n,n} - N_{n-i,n-i'}
	\;\raa\; (N_{j,j'} + N_{n-j,n-j'}) - N_{n-i,n-i'}\]
for each $0 \leq i, i', j, j' \leq n$.
~
That is, each nonterminal $N_{j,j'}$ occurs in a derivation from a
start symbol $N_{i,i'}$.
\qed
}

In these cases, in order to find a minimal weight term for
the goal nonterminal
$N_{g_1,g_2}$, we have to consider $N_{j,j'}$ for every possible
combination $j,j'$, anyway.
~
That is, considering only nonterminals that
are reachable from $N_{g_1,g_2}$ in $\G_{1,2}$
doesn't restrict the search space.
~
Hence, we do not need to consider a grammar at all;
we merely have to apply the given operations $f$ and build term normal
forms.
~
More precisely, step~\ref{knuth apply}
of Alg.~\ref{Knuth's algorithm for grammar schemes}
can be replaced by just building $f( \:,1k{\phi(N_{i_\i,i'_\i})})$
as soon as all $\:,1k{\phi(N_{i_\i,i'_\i})}$ are defined,
and considering it as a possibly minimal term for
$N_{\nf(f(\:,1k{i_\i})), \nf(f(\:,1k{i'_\i}))}$.
~
Since we don't need grammars or grammar schemes any longer, we don't
have to bother about justifying our above optimistic assumptions about
implementability of the latter.

Knuth original algorithm is just about minimal weights, not minimal
weight terms;
it uses a heap to find a minimal weight
among the considered weights.
~
When dealing with weight terms instead,
collecting all terms
corresponding to a \notion{weight decomposition list} is an
optimization.
~
When we implement it, we arrive
at Alg.~\ref{Value-pair cached term generation}.
~
Thus, the latter has been obtained as an improvement of the
Constrained E-generalization algorithm
from \cite{Burghardt.2005c}.

From an implementation point of view, the main advantage of
Alg.~\ref{Value-pair cached term generation}
is that it doesn't need to store huge grammars in memory.
~
Starting e.g.\ from a
grammar $\G$ for the equivalence classes of
$0,\ldots,120$ w.r.t.\ just $+$ and $-$,
the
constrained $E$-generalization algorithm
from \cite{Burghardt.2005c}
arrives at a grammar $\G_{1,2}$ with $121^2 = 14641$
rules and a total of more than
$\sum_{n,n'=0}^{120} (n\.+1)(n'\.+1) + (121\.-n)(121\.-n')
= 108\,958\,322$
alternatives.
~
$\G_{1,2,3}$ and $\G_{1,2,3,4}$ have
already more than
$8 \cdot 10^{11}$
and
$5 \cdot 10^{15}$
alternatives, respectively.
~
Figure~\ref{Grammar size vs. tuple length for 0,...,120 and +,-}
summarizes these relations.
~
In this example, the alternative count is of magnitude of the
nonterminal count's square.
~
The Constrained E-generalization algorithm from
\cite{Burghardt.2005c} needs memory proportional to the
alternatives count, while both $\D$ and $\phi$ in 
Alg.~\ref{Term generation from decomposition lists} need memory
proportional to the nonterminal count.
~
The size of $\F$ in Alg.~\ref{Weight decomposition list generation}
is bounded by $w^a$ where $w$ is the number of distinct weights and
$a$ the maximal arity of operators; this bound can't get larger than
the alternatives count, and usually is much smaller.

\begin{figure}
\newcommand{\0}[2]{#2}
\begin{center}
\renewcommand{\arraystretch}{1.0}
$\begin{array}{|l||r|r|r|r|}
\hline
\mbox{Tuple length} & 1 & 2 & 3 & 4	\\
\hline
\mbox{Nonterminals} 
	& \0{1 \cdot 10^2}{121}
	& \0{1 \cdot 10^4}{14\,641}
	& \0{1 \cdot 10^6}{1\,771\,561}
	& \0{2 \cdot 10^8}{214\,358\,881}
	\\
\mbox{Alternatives}
	& \0{1 \cdot 10^4}{14\,762}
	& \0{1 \cdot 10^8}{108\,958\,322}
	& \0{8 \cdot 10^{11}}{804\,221\,374\,682}
	& \0{5 \cdot 10^{15}}{5\,935\,957\,966\,527\,842}
	\\
\hline
\end{array}$
\caption{Grammar size vs. tuple length for $0,\ldots,120$ and $+$,$-$}
\LABEL{Grammar size vs. tuple length for 0,...,120 and +,-}
\end{center}
\end{figure}

The IST'70-ZR intelligence test could not be tackled with an algorithm
that needs to store grammars corresponding to product automata,
even with today's main memory technology.
~
(It can be tackled, however, with the implementation of
Alg.~\ref{Value-pair cached term generation},
a report is forthcoming.)

$$*$$

It is instructive to compare
the best algorithm we could come up with, i.e.\
Alg.~\ref{Value-pair cached term generation},
with the most naive algorithm for the same task.
~
The latter simply generates all terms in order of increasing weight,
and stops as soon as $t \sigma_1 =_E g_1$ and $t \sigma_2 =_E g_2$
for the current term $t$.
~
The only difference to Alg.~\ref{Value-pair cached term generation}
is that it doesn't cache the minimal weight terms for each value
combination.

Arriving, after proceeding on a long and sophisticated path, that
close to the starting point, is rather discouraging.
~
It may indicate that there is no better way to solve the task,
at least not on a Turing / Von Neuman architecture.
~
A possible remedy might be to devise
a highly parallel architecture to
synthesize nonrecursive function definition terms from examples.
~
This would comply with the neuronal structure of a human brain.

\EXAMPLE{Interleaved sequence}{%
\LABEL{Interleaved sequence}%
As a closing example for this section, consider the sequence
$0, 1, 2, 1, 4, 1, 6, 1, 8, 1, \ldots$.
~
Its most obvious construction law term is
${\it if}(v_p \% 2 = 0, v_p, 1)$,
where ${\it if}(x,y,z)$ and $x \% y$ denotes a case distinction 
and the integer
remainder, written in the {\tt C} programming language as ``{\tt
(x?y:z)}'' and ``{\tt x\%y}'', respectively.
~
The Lagrange interpolation term for this sequence is
$$
  \frac{-1}{405} v_p^9 
+ \frac{31}{315} v_p^8   
- \frac{1556}{945} v_p^7   
+ \frac{676}{45} v_p^6 
- \frac{2192}{27} v_p^5   
+ \frac{11872}{45} v_p^4   
- \frac{201536}{405} v_p^3   
+ \frac{17152}{35} v_p^2   
- \frac{59077}{315} v_p
,$$
it has the size $116$ even if each constant is considered to be 
only of size $1$.
~
Due to the huge number of value-tuples resulting from
terms of smaller size, our implementation is unable to find the
Lagrange term.

However, if the set of available operators in law terms is varied,
surprisingly unexpected law terms are found.
~
Some of them are shown in
Fig.~\ref{Law terms in Exm. Interleaved sequence},
where a column below an operator shows its evaluation result on the
sequence, and
a column corresponding to a law term's root is shown in
boldface.
~
We denote by $\undef$, $\false$, and $\true$ an undefined value, the
boolean falsity, and truth, respectively.
~
We use $/$ and $\idiv$ to denote the ordinary and the truncating integer
division, for example, $7/2$ and $7 \idiv 2$ 
yields $\undef$ and $3$, respectively.

We could find neither a law term built from only $\%, v_p, v_1$
nor one built from only $+, v_p, v_1$; 
see Fig.~\ref{Failed attempts in Exm. Interleaved sequence} 
for details.
\qed
}

\begin{figure}
{
\newcommand{\0}{\scriptstyle}
\newcommand{\1}{\mathbf}
\renewcommand{\arraystretch}{1.0}
$$\begin{array}{| c*{6}{@{}c@{}}c | c*{7}{@{}c@{}}c | c*{5}{@{}c@{}}c
| c*{5}{@{}c@{}}c | c@{}c@{}c |}
\hline
{\it if} & (v_p & \% & 2 & =      & 0, & v_p, & 1)      & v_p &    - & (v_p & \% & 2) & * & (v_p & -      & 1)      & v_p &    - & (v_1 & \idiv & 2) & * & 2      & 2 &  * & (v_p & - & v_1) &    - & v_2      & v_p  & \idiv  & v_1	\\ \hline 
\1     0 & \0 0 &  0 &   & \true  &    & \0 0 &         &\0 0 & \1 0 & \0 0 &  0 &    & 0 & \0 0 & \undef &         &     &      &      &      &    &   &        &   &    &      &   &      &      &          &      &       &    	\\
\1     1 & \0 1 &  1 &   & \false &    & \0 1 &         &\0 1 & \1 1 & \0 1 &  1 &    & 0 & \0 1 & 0      &         &\0 1 & \1 1 & \0 0 &    0 &    & 0 &        &   &    &      &   &      &      &          &      &       &    	\\
\1     2 & \0 2 &  0 &   & \true  &    & \0 2 &         &\0 2 & \1 2 & \0 2 &  0 &    & 0 & \0 2 & 1      &         &\0 2 & \1 2 & \0 1 &    0 &    & 0 &        &   &  2 & \0 2 & 1 & \0 1 & \1 2 &\0 0      &\0 2  & \1  2 &\0 1	\\
\1     1 & \0 3 &  1 &   & \false &    & \0 3 &         &\0 3 & \1 1 & \0 3 &  1 &    & 2 & \0 3 & 2      &         &\0 3 & \1 1 & \0 2 &    1 &    & 2 &        &   &  2 & \0 3 & 1 & \0 2 & \1 1 &\0 1      &\0 3  & \1  1 &\0 2	\\
\1     4 & \0 4 &  0 &   & \true  &    & \0 4 &         &\0 4 & \1 4 & \0 4 &  0 &    & 0 & \0 4 & 3      &         &\0 4 & \1 4 & \0 1 &    0 &    & 0 &        &   &  6 & \0 4 & 3 & \0 1 & \1 4 &\0 2      &\0 4  & \1  4 &\0 1	\\
\1     1 & \0 5 &  1 &   & \false &    & \0 5 &         &\0 5 & \1 1 & \0 5 &  1 &    & 4 & \0 5 & 4      &         &\0 5 & \1 1 & \0 4 &    2 &    & 4 &        &   &  2 & \0 5 & 1 & \0 4 & \1 1 &\0 1      &\0 5  & \1  1 &\0 4	\\
\1     6 & \0 6 &  0 &   & \true  &    & \0 6 &         &\0 6 & \1 6 & \0 6 &  0 &    & 0 & \0 6 & 5      &         &\0 6 & \1 6 & \0 1 &    0 &    & 0 &        &   & 10 & \0 6 & 5 & \0 1 & \1 6 &\0 4      &\0 6  & \1  6 &\0 1	\\
\1     1 & \0 7 &  1 &   & \false &    & \0 7 &         &\0 7 & \1 1 & \0 7 &  1 &    & 6 & \0 7 & 6      &         &\0 7 & \1 1 & \0 6 &    3 &    & 6 &        &   &  2 & \0 7 & 1 & \0 6 & \1 1 &\0 1      &\0 7  & \1  1 &\0 6	\\
\1     8 & \0 8 &  0 &   & \true  &    & \0 8 &         &\0 8 & \1 8 & \0 8 &  0 &    & 0 & \0 8 & 7      &         &\0 8 & \1 8 & \0 1 &    0 &    & 0 &        &   & 14 & \0 8 & 7 & \0 1 & \1 8 &\0 6      &\0 8  & \1  8 &\0 1	\\
\1     1 & \0 9 &  1 &   & \false &    & \0 9 &         &\0 9 & \1 1 & \0 9 &  1 &    & 8 & \0 9 & 8      &         &\0 9 & \1 1 & \0 8 &    4 &    & 8 &        &   &  2 & \0 9 & 1 & \0 8 & \1 1 &\0 1      &\0 9  & \1  1 &\0 8	\\ \hline
\end{array}$$
}
\caption{Law terms in Exm.~\ref{Interleaved sequence}}
\LABEL{Law terms in Exm. Interleaved sequence}
\end{figure}

\begin{figure}[b]
\begin{center}
\renewcommand{\arraystretch}{1.0}
\begin{tabular}{|l|rrrrr|}
\hline
Optr set & Sz & Terms & Computations & Memory (bytes) & Time (sec)	\\
\hline
$\%, v_p, v_1$ & $36$ & $9688$ & $51979260$ & $134990875$ & $11$	\\
$+, v_p, v_1$ & $726$ & $243035$ & $19408297875$ & $159683896$ & $3218$	\\
\hline
\end{tabular}
\caption{Failed attempts in Exm.~\ref{Interleaved sequence}}
\LABEL{Failed attempts in Exm. Interleaved sequence}
\end{center}
\end{figure}

\clearpage
\section{Implementation overview}
\LABEL{Implementation overview}

In this section, we sketch some features of our \code{C}
implementation.
~
Section~\ref{Kernel modules overview} gives a brief overview of all
kernel modules,
while Sect.~\ref{User modules overview}
sketches the modules that hold user-definable code.

The subsequent sections elaborate on certain approaches to optimize
the term-generation process.
~
We discuss
\begin{itemize}
\item
	in Sect.~\ref{Pruning for binary operators}
	how to avoid building redundant
	commutative and associative variants of terms, 
	like $(t_2+t_1)+t_3$
	when $t_1+(t_2+t_3)$ already was built;
\item
	in Sect.~\ref{Module redex.c} 
	how to avoid building other redundant terms,
	like $t-t$;
\item
	in Sect.~\ref{Many-sorted signatures for operators}
	how to avoid building nonsensical terms,
	like $(t_1 \leq t_2)*t_3$;
\item
	in Sect.~\ref{Projection-like operators}
	how to efficiently build terms involving
	${\it if}(\cdot,\cdot,\cdot)$
	and similar operators.
\item
	in Sect.~\ref{Module assoc.c}, how to implement the mapping
	$\phi$ from
	Alg.~\ref{Term generation from decomposition lists}
	efficiently.
\end{itemize}

\subsection{Kernel modules overview}
\LABEL{Kernel modules overview}

We give a short description of each module file of the
\code{C}-implementation:

\begin{enumerate}
\item Identifiers and syntax

	\begin{enumerate}
	\item\LABEL{stringTab.c}
		Module \code{stringTab.c}
		--- Administration of strings.
	\item\LABEL{idTab.c}
		Module \code{idTab.c}
		--- Administration of operator and variable identifiers.
		~
		This module implements the 
		scanner underlying the parser in
		module~\ref{parser.c}.
	\item\LABEL{parser.c}
		Module \code{parser.c}
		--- Parsing routines:
		\code{parseValues},
		\code{parseOpDefs},
		\code{parse\-Redices},
		\code{parseVariables},
		\code{parseGoals}
	\item\LABEL{wgfTab.c}
		Module \code{wgfTab.c}
		--- Administration of weight functions (like \code{size},
		\code{height}); each of those functions is made
		are accessible by a name.
		~
		The actual code of weight functions is contained in
		module \code{UserWgf.c} 
		(module~\ref{UserWgf.c} 
		in Sect.~\ref{User modules overview}).
	\end{enumerate}
\item Values, sorts, and redices
	\begin{enumerate}
	\item\LABEL{valDefTab.c}
		Module \code{valDefTab.c}
		--- Administration of normal forms (``values'').
		~
		While from a theoretical viewpoint it is convenient to
		consider values to be terms in normal form
		(Def.~\ref{Equality, normal form},
		our implementation represents each value by an
		integer.
		~
		Depending on the sort (see module~\ref{opTab.c}) it
		belongs to, it may be interpreted as an index into
		into an identifier table \code{valTab[]}
		(like \code{"true"},
		\code{"false"} for $\true$ and $\false$, respectively),
		or handled by user-defined scan and print routines.
		~
		This module provides access functions to
		\code{valTab[]}.
	\item\LABEL{opTab.c}
		Module \code{opTab.c}
		--- Administrates user-defined sorts.
		~
		Allows for building a new sort from a given set of values.
		~
		Checks user-claimed algebraic properties (associativity,
		commutativity, idempotency) of operator functions;
		this is feasible since all sorts are finite; to save
		computation time, only border-cases may be checked for
		associativity.
		~
		Assigns an inhabited\footnote{%
			at least one term of this sort can be built%
		}%
		- and needed\footnote{%
			at least one term of the sort of a goal
			contains a subterm of this sort%
		}%
		-flag to each sort.
	\item\LABEL{opDefTab.c}
		Module \code{opDefTab.c}
		--- Administrates user-defined operators and functors,
		including variables.
		~
		The execution routine of such operators and functors
		simply does table-lookup in \code{opDefTab[]}.
		~
		Operator and functor result values are read by
		\code{parseOpDefs} and
		stored
		consecutively, least argument running fastest, first
		one running slowest.

	\item\LABEL{redex.c}
		Module \code{redex.c}
		--- Administration of user-provided redices.
		~
		See Sect.~\ref{Module redex.c}.
	\item\LABEL{sorts.c}
		Module \code{sorts.c}
		--- Operator selection from argument or result sort.
		~
		See Sect.~\ref{Many-sorted signatures for operators}.

	\end{enumerate}
\item Goals and solution terms
	\begin{enumerate}
	\item\LABEL{term.c}
		Module \code{term.c}
		--- Generic term traversal routines;
		term enumeration;
		term normal-form computation
		(i.e.\ evaluation using the user-defined operator
		implementation routines)
	\item\LABEL{contTab.c}
		Module \code{contTab.c}
		--- Administrates the set $\D$ of terms of minimal
		weight for their result vector.
		~
		Those terms are ordered by weight and main operator.
		~
		The table
		\code{cont\-Chain\-Start[]}
		allows one to access a list of all terms of a given weight.
		~
		The function
		\code{idIT NextInhbArgOp(maIT *maList,cnIT *cnChain);}
		allows one to find the next operator common to
		\code{maList} (see
		module~\ref{sorts.c}) and \code{cnChain}.

	\item\LABEL{goals.c}
		Module \code{goals.c}
		--- Administrates given goal vectors.
		~
		Allows for multiple goals.
		~
		Allows the \code{compute} algorithm
		(module~\ref{compute.c}) to
		enter a preliminary (non-minimal) as well as
		a final (minimal) solution term for a goal.
		~
		Implements a timeout-mechanism to stop search and
		return the
		best (i.e.\ least-weight) solution terms found so far.
	\end{enumerate}
\item Overall control

\begin{figure}
\begin{center}
\renewcommand{\arraystretch}{1.0}
\begin{tabular}{|l|ll|}
\hline
\code{"val"}   & \multicolumn{2}{l|}{obtain the set of values} \\
& \code{"parse"} & read from file \\
\hline
\code{"def"}   & \multicolumn{2}{l|}{obtain operator functionality} \\
& \code{"parse"} & read from file \\
\hline
\code{"red"}   & \multicolumn{2}{l|}{obtain redices 
	(Sect.~\ref{Module redex.c})} \\
& \code{"parse"} & read from file \\
\hline
\code{"var"}
	& \multicolumn{2}{l|}{obtain variables
	and their value vectors} \\
& \code{"parse"} & read from file \\
& \code{"all"} & one variable for each combination of values \\
& \code{"seq"} & generate from given value sequence \\
& \code{"univ"} & generate universal substitution variables \\
\hline
\code{"goals"} & \multicolumn{2}{l|}{obtain goals
	and their value vectors} \\
& \code{"parse"} & read from file \\
& \code{"seq"} & generate from given value sequence \\
\hline
\code{"upper"}
	& \multicolumn{2}{l|}{obtain settings for projection-like operators
	(Sect.~\ref{Projection-like operators})} \\
& \code{"none"} & do not use any \\
& \code{"idx"} & use array indexing \\
& \code{"condChoice"} & use kind of ${\it if}(\cdot,\cdot,\cdot)$ \\
\hline
\end{tabular}
\caption{Execution phases / contexts in module~\ref{param.c}}
\LABEL{Execution phases / contexts fig}
\end{center}
\end{figure}

	\begin{enumerate}
	\item\LABEL{param.c}
		Module \code{param.c}
		--- Administration of command line arguments and
		global variables.
		~
		As a poor-man's substitute for object oriented programming%
			\footnote{%
			In fact, all but the core algorithm should
			better have
			been implemented in \code{C++}.%
			}%
		, we separated the program execution into phases%
			\footnote{%
			called ``contexts'' in the C source code,
			in order to avoid confusion with a lower-level
			notion of ``phase'' used there for debugging
			purposes only%
			}%
		, and implemented the execution of each phase by an
		indirect call only.
		~
		This allowed us to change the behavior of each phase
		by command line options.
		~
		Figure~\ref{Execution phases / contexts fig}
		shows a list of the execution phases.
		~
		See also module~\ref{main.c}.
	\item\LABEL{main.c}
		Module \code{main.c}
		--- As described in module~\ref{param.c}, \code{main()}
		just calls for each phase
		the corresponding routine, as set by the command
		line arguments.
	\end{enumerate}
\item Core algorithm
	\begin{enumerate}
	\item\LABEL{bbt.c}
		Modules \code{bbtTab.c} and \code{bbt.c}
		--- Implements a balanced binary tree; keys are indices
		into \code{contTab} (module~\ref{contTab.c}).
		~
		To save memory, we don't use nodes
		without child pointers; the data field of such a node
		is instead stored directly in the pointer field of its
		parent node.
	\item\LABEL{assoc.c}
		Module \code{assoc.c}
		--- Implements a lookup mechanism for indices into
		\code{contTab} on top of module~\ref{bbt.c}.
		~
		Discussed in detail in Sect.~\ref{Module assoc.c}.
	\item\LABEL{heap.c}
		Module \code{heap.c}
		--- Implements a double-ended
		priority queue of weight-term
		indices, ordered by result weight; it is possible
		to extract the set of all weight-terms with least
		weight simultaneously.
		~
		In case of imminent memory overflow, the largest
		weights may be efficiently found and discarded.
	\item\LABEL{weightTerm.c}
		Module \code{weightTerm.c}
		--- Administration of weight decomposition lists
		(called ``weight-terms'' in the implementation);
		each one can be accessed by an index of type \code{wtIT}
		that points to the \code{wtTab[]} table defined in
		this module.
	\item Module \code{computeWt.c}
		--- Implements
		Alg.~\ref{Weight decomposition list generation}.
		~
		Provides optimized routines for commutative operators,
		see Sect.~\ref{Pruning for binary operators}.
	\item\LABEL{compute.c}
		Module \code{compute.c}
		--- Provides the \code{compute} routine,
		the implementation of
		Alg.~\ref{Term generation from decomposition lists}
		and~\ref{Value-pair cached term generation}.
		~
		It includes optimized routines for associative,
		commutative, or idempotent operators,
		see Sect.~\ref{Pruning for binary operators}.
	\end{enumerate}
\item\LABEL{upper.c}
	Module \code{upper.c}
	---
	Administration of projection-like operators, 
	see Sect.~\ref{Projection-like operators}
	(``upper'' refers to their position at the terms' roots).
\item Equation generation
	--- Used to generate equations from a given finite algebra,
	as described in \cite{Burghardt.2002b}.
	~
	Not completely implemented, and not described here.
\item Module \code{User.c}
	--- Interface to user-definable code 
	(see Sect.~\ref{User modules overview});
	contains initialization and finalization routines.
\end{enumerate}

\subsection{User modules overview}
\LABEL{User modules overview}

As a naming convention, file names starting with ``\code{User}''
indicate user-definable code.
~
Whenever nontrivial parametrizations are to be user-provided
we followed the approach to include them into the \code{C} source
core, instead of providing an own parametrization language,
this way saving the effort of implementing an according interpreter.
~
Therefore, the user is required to re-compile the whole software e.g.\
after adding another weight function routine to module
\code{UserWgf.c}.
~
We briefly sketch each user module in the following.

\begin{enumerate}
\item Module \code{UserOp.c}
	--- Interface to operator-defining user modules:
	~
	For each operator, we have to provide a computation routine,
	an initialization routine (called whenever the operator of the
	weight decomposition list changes in 
	Alg.\ref{Term generation from decomposition lists}'s
	input), 
	property flags, and a brief textual description to be
	shown by the command-line option \code{-help-ops}.
	~
	Operators are grouped by application domain:
	\begin{enumerate}
	\item Module \code{UserOpArith.c}
		--- Provides user-defined arithmetic operators, like
		$+$, $-$, $*$, $/$, $/\!/$, $\%$, $\min$, $\max$,
		$<$, $\leq$, $=$, $\geq$, $>$,
		$\lnot$, $\land$, $\lor$, $\Rightarrow$
	\item Module \code{UserOpBit.c}
		--- Provides bitwise operators on unsigned integers, like
		\code{\&}, \code{|}, 
		\code{\^{}},
		\code{\~{}}, 
		\code{<<}, \code {>>}
	\item Module \code{UserOpString.c}
		--- String operations (experimental, only a tiny
		character set and short maximal length can be
		supported), like
		concatenation, reversal, head, tail, character
		replacement, rotation, interleaving, length.
	\end{enumerate}
\item Module \code{UserTerm.c} (doesn't work yet)
	--- symbolic terms as values.
	\begin{enumerate}
	\item Module \code{UserTermArith0s.c} (doesn't work yet)
		--- example application to symbolic terms built from 
		\code{0} and \code{succ},
		shall also eventually provide \code{+} and \code{*}.
	\item Module \code{UserTermLambda.c} (doesn't work yet)
		--- symbolic typed lambda terms.
	\item Module \code{UserTermList.c} (doesn't work yet)
		--- symbolic list built from
		\code{nil} and \code{cons}.
	\end{enumerate}
\item Module \code{UserUp.c}
	--- Interface to projection-like operator modules, 
	see Sect.~\ref{Projection-like operators} for details.
	~
	The following operator kinds are provided:
	\begin{enumerate}
	\item Module \code{UserUpCondChoice.c}
		--- defines 
		${\it if}(\cdot,\cdot,\undef)$
		and an angelic nondeterministic choice.
	\item Module \code{UserUpIdx.c}
		--- defines operators to select from a given array.
	\item Module \code{UserUpNone.c}
		--- trivial implementations to be used by the kernel
		code in absence of any projection-like operator.
	\end{enumerate}
\item\LABEL{UserVal.c}
	Module \code{UserVal.c}
	--- defines routines to scan and print used-defined values,
	including (experimental) symbolic terms.
\item Variable definitions modules:
	\begin{enumerate}
	\item Module \code{UserVarAll.c}
		--- code to generate one variable for every possible
		combination of values, or for random combination of
		values.
	\item Module \code{UserVarUniv.c}
		--- code to generate variables corresponding to 
		``universal substitutions'', as described in 
		\cite[Lem.5, p.7]{Burghardt.2005c}.
	\item Module \code{UserSeq.c}
		--- Application of E-anti-unification to sequence law
		guessing, as described in 
		\cite[Sect.5.2, p.28--29]{Burghardt.2005c}.
	\end{enumerate}
\item\LABEL{UserWgf.c}
	Module \code{UserWgf.c}
	--- define code for weight functions, like
	\code{size}, and \code{height}.

\end{enumerate}

Next, we elaborate on some selected details of our implementation.

\newcommand{\ts}{{\sf ts}}

\subsection{Pruning for binary operators}
\LABEL{Pruning for binary operators}

When a binary operator $f \in \Sigma$ is associative, commutative, or
idempotent, it is reasonable to expect that
its weight function $\wg{f}: \W \times \W \raa \W$
shares the same properties.
~
While we don't require this correspondence, we provide optimizations
of Alg.~\ref{Value-pair cached term generation}
for those operators that obey it.

For this purpose, the following total well-ordering $(\succeq)$ on
terms has been approved useful:
define
$f(\:,1n{t_\i}) \succeq f'(\:,1{n'}{t'_\i})$
iff
\[
\begin{array}{lll}
\WG(f(\:,1n{t_\i})) > \WG(f'(\:,1n{t'_\i})),
	&& \mbox{ or}	\\
\WG(f(\:,1n{t_\i})) = \WG(f'(\:,1n{t'_\i}))
	& \land
	& \ts(f) \geq \ts(f'),\\
\end{array}
\]
where $\ts: \Sigma \raa \N$
is an arbitrary ranking function\footnote{
	Its name derives from ``time stamp'', as we use just the
	order in which function symbols are entered into the symbol
	table in
	module~\ref{idTab.c} in Sect.~\ref{Kernel modules overview}.
}
on the set of operator symbols.
~
In particular, $t_1 \succeq t_2$ implies $\WG(t_1) \geq \WG(t_2)$ for
arbitrary terms $t_1, t_2$, so the former ordering is a refinement of
the latter.

In Alg.~\ref{Weight decomposition list generation}, we order the
weight decomposition lists in the $\F$ heap in fact by $\succeq$,
rather than just by weight.
~
More precisely, we compare such lists by
\[
\begin{array}{r rclll}
& \tpl{f,\:,1n{w_\i}} & \succeq & \tpl{f',\:,1n{w'_\i}}	\\
\mbox{ iff }
	& \wg{f}(\:,1n{w_\i}) & > & \wg{f}'(\:,1n{w'_\i})	\\
\mbox{ or }
	& \wg{f}(\:,1n{w_\i}) & = & \wg{f}'(\:,1n{w'_\i})
	& \land 
	& \ts(f) \geq \ts(f').	\\
\end{array}
\]
As a consequence
in Alg.~\ref{Term generation from decomposition lists},
the set $\D$ holding the minimal terms generated so far is segmented not
just by increasing weight,
but moreover by increasing $\ts(\cdot)$ rank of
the main operator.
~
That is, for each $f \in \Sigma$, the set
\[
\D_{w,f} =
\set{ f( \:,1n{t_\i} ) \in \D_w \mid \:,1n{t_\i} \in \T{}{}{} }
\]
corresponds to a continuous segment of the list implementing
$\D_w$.
~
The finer segmentation is indicated by light grey bars in 
Fig.~\ref{Example state in Alg. (red) and (blue)}, assuming
$\ts(0) < \ts(1) < \ts(2)
< \ts(v_p) < \ts(v_1) < \ts(v_2)
< \ts(+) < \ts(*)$.

If some operator $\oplus$ as well as its weight function $\wg{\oplus}$
is known to be commutative, we
need to build the term $t_1 \oplus t_2$ in step~\ref{cached term-gen build}
of Alg.~\ref{Term generation from decomposition lists}
only if $t_1 \succeq t_2$.
~
The term $t_2 \oplus t_1$ needn't be built in that case since
it evaluates to the same value pair and has
the same weight as $t_1 \oplus t_2$.
~
Since building the former term is about as fast as testing the latter
condition, we can't save time just by guarding the former by the
latter.
~
Instead, we need to restrict the
loop~\ref{cached term-gen loop over terms}
in Alg.~\ref{Term generation from decomposition lists}
such that it ranges only over $t_1 \in \D_{w_1}, t_2 \in D_{w_2}$
with $t_1 \succeq t_2$.

To this end, we insert in step~\ref{Weight-gen heap insert}
of Alg.~\ref{Weight decomposition list generation}
a weight decomposition list $\tpl{\oplus,x_1,x_2}$
into $\F$ only if $x_1 \geq x_2$.
~
Then, in step~\ref{cached term-gen draw}
of Alg.~\ref{Term generation from decomposition lists},
only these lists will be read for $\oplus$,
and hence the loop~\ref{cached term-gen loop over terms}
will range only over terms $t_1, t_2$ of weight $w_1, w_2$
with $w_1 \geq w_2$.
~
Since this covers all $t_1, t_2$ with $t_1 \succeq t_2$,
the algorithm remains complete.
~
It saves execution time since we have, per weight decomposition list
$\tpl{\oplus,x_1,x_2}$,
one extra test $x_1 \geq x_2$
in Alg.~\ref{Weight decomposition list generation},
but avoid building
in Alg.~\ref{Term generation from decomposition lists}
all terms from the usually large set
$\set{ t_1 \oplus t_2 \mid t_1 \in D_{x_1}, t_2 \in D_{x_2}}$
when the test fails.

Similarly,
if $\oplus$ and $\wg{\oplus}$
is known to be both associative and commutative,
we don't need to build
terms of the form $t'_1 \oplus (t'_2 \oplus t'_3)$ at all,
and need to build $(t'_1 \oplus t'_2) \oplus t'_3$
only if $t'_1 \succeq t'_2 \succeq t'_3$.
~
Again, due to $\oplus$'s commutativity,
it is sufficient in step~\ref{Weight-gen heap insert}
of Alg.~\ref{Weight decomposition list generation}
to insert a weight decomposition list $\tpl{\oplus,x_1,x_2}$
into $\F$ only if $x_1 \geq x_2$.
~
Moreover,
weight decomposition lists of the form $\tpl{\oplus,x_1,x_2}$
are handled differently in the
steps below~\ref{cached term-gen empty}
of Alg.~\ref{Term generation from decomposition lists}.
~
If, in step~\ref{cached term-gen loop over terms},
$t_1$ is currently from $\D_{w_1,\oplus}$,
i.e.\ if has the form $t_1 = t'_1 \oplus t'_2$,
we choose $t_2$ such that $t'_2 \succeq t_2$,
i.e.\ we choose $t_2$ from 
\[
\bigcup 
\set{\D_{w_2,f} \mid w_2 \leq \WG(t'_2) \land \ts(f) \leq \ts(\oplus)}.
\]
This is achieved by function \code{selectArgOp1\_2AC} in file
\code{compute.c}.

\begin{figure}
\begin{center}
\includegraphics[width=\textwidth]{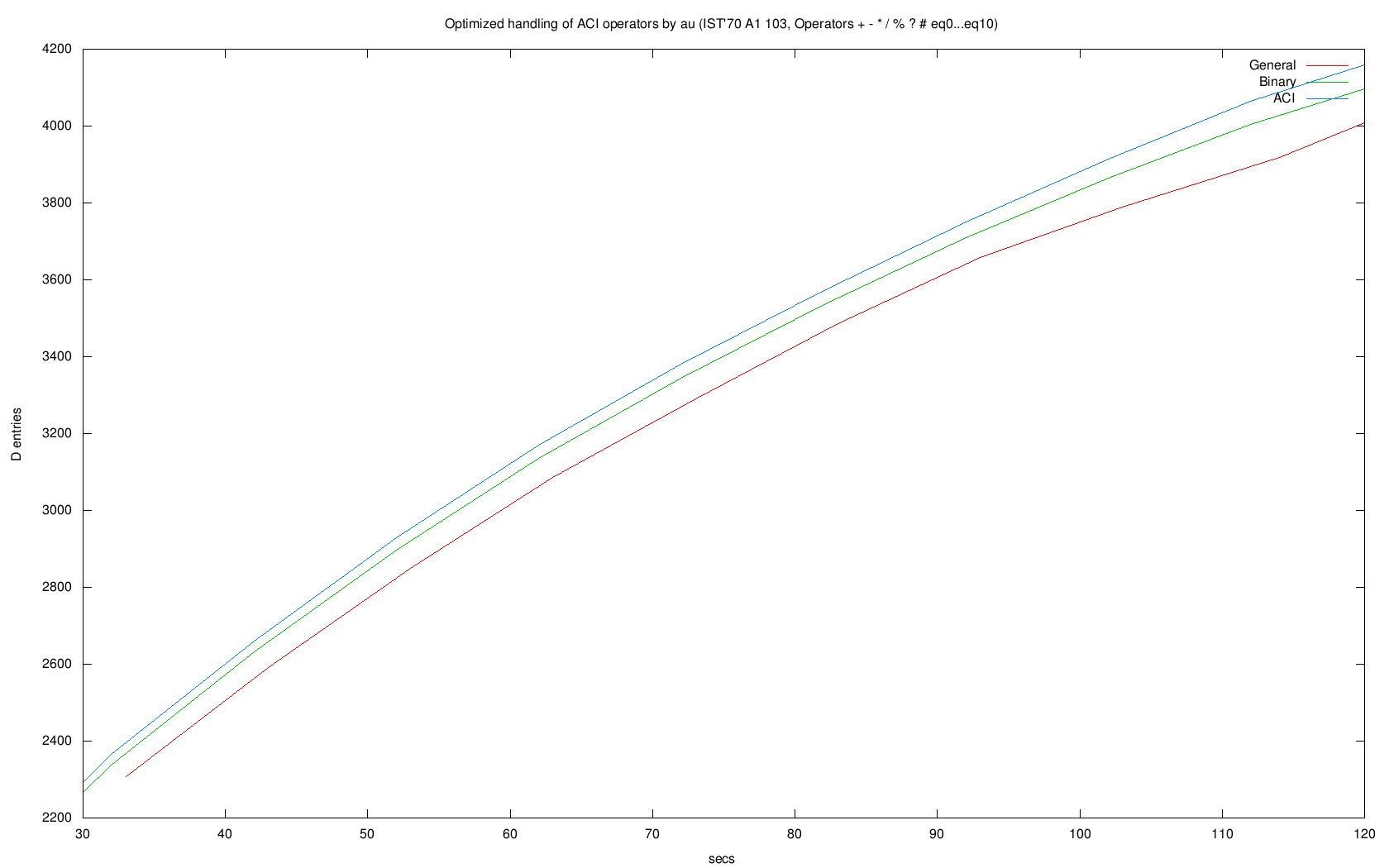}
\caption{Acceleration by ACI pruning}
\label{Acceleration by ACI pruning}
\end{center}
\end{figure}

Figure~\ref{Acceleration by ACI pruning}
shows the effect of this kind of pruning for associative, commutative,
or/and idempotent operators on a practical example.\footnote{
	Attempt to compute, from the operators listed in the caption,
	a law term for the integer sequence
	$55, 57; 60, 20, 10, 12, 15$
	(task ``A1 103'' of the intelligence test IST'70-ZR).
}
~
It shows the growth of $\D$ vs.\ the computation time for three different
implementations of Alg.~\ref{Value-pair cached term generation}:
\begin{itemize}
\item ``{\em General\/}'' --- without optimization
\item ``{\em Binary\/}'' --- uses particularized subroutines in
	Alg.~\ref{Term generation from decomposition lists}
	whenever a weight decomposition list with a binary operator is
	handled
\item ``{\em ACI\/}'' --- additionally implements pruning for
	associative, commutative, and/or idempotent operators as
	described above
\end{itemize}
The data shows that the improvement by the latter is rather small;
after 120 seconds of computation we have a ratio 
{\em ACI}:{\em Binary\/} = 101.5\% and 
{\em ACI}:{\em General\/} = 103.8\%.

\subsection{Module \code{redex.c}}
\LABEL{Module redex.c}

This module handles user-provided redices.
~
For example, terms of the form $x*(y+z)$ need not be generated
in addition to terms of the form $(x*y)+(x*z)$ if $*$ is known to
distribute over $+$.

Only linear terms\footnote{
	i.e.\ terms without multiple occurrences of a variable
}
of depth 2 can be given as redices;
other terms, like e.g.\
$x-x$ and $\sqrt{x^2}$, cannot be handled.
~
Redices are given in the form
\code{MainOp ArgOp1 \ldots\ ArgOpN},
where \code{N} is the arity of \code{MainOp},
meaning that term of the form
$MainOp(ArgOp1(\ldots),\ldots,ArgOpN(\ldots))$
need not be built.
~
A single dot ``\code{.}'' denotes a fresh variable.
~
In the above distributivity
example, the redex ``\code{* . +}'' would be provided,
denoting the redex $x_1 * (x_{21} + x_{22})$.

Simple redices, i.e.\ redices where \code{ArgOpI} is
``\code{.}''
for all but one value of \code{I}, are considered in module
\code{sorts.c} (see Sect.~\ref{Many-sorted signatures for operators});
they need not be kept in the
redex table.
~
All remaining redices are kept in that table, indexed by their
main operator, and are used in the routine
\code{bool isRedex(idIT mainOp,const idIT argOp[])}.

Filtering based on \code{isRedex} is done after drawing a weight
term, and after choosing \code{argOp}s, but before choosing terms from $\D$,
in the routines \code{selectDnArgOpsV}, \code{selectArgOp1\_2},
\code{select\-Arg\-Op1\_2C}, \code{selectArgOp1\_2AC}, \code{selectWt1},
all from file \code{compute.c}.

\subsection{Many-sorted signatures for operators}
\LABEL{Many-sorted signatures for operators}

\begin{figure}
\renewcommand{\arraystretch}{1.0}
\begin{tabular}[t]{|r@{}l@{$\;$}l@{$\;$}l@{$\;$}l@{$\;$}l|l}
\cline{1-6}
$0$ & : &       &      & $\raa$ & int	\\
$v$ & : &       &      & $\raa$ & int	\\
$+$ & : & int,  & int  & $\raa$ & int   & c,a	\\
$-$ & : & int,  & int  & $\raa$ & int	\\
$*$ & : & int,  & int  & $\raa$ & int   & c,a	\\
$<$ & : & int,  & int  & $\raa$ & bool	\\
$\lor$  & : & bool, & bool & $\raa$ & bool  & c,a,i	\\
$\land$ & : & bool, & bool & $\raa$ & bool  & c,a,i	\\
$\lnot$ & : & bool  &      & $\raa$ & bool	\\
\cline{1-6}
\end{tabular}
\hfill
\begin{tabular}[t]{|lr@{$\;$}l|l@{$\;$}l}
\cline{1-3}
$+$ & $0$ & .   & \multicolumn{2}{l}{not useful for pruning}	\\
$+$ & .   & $0$ & \multicolumn{2}{l}{not useful for pruning}	\\
$*$ & $+$ & .   & $(a + b) * c$    & $= a * c + b * c$	\\
$*$ & .   & $+$ & $a * (b + c)$    & $= a * b + a * c$	\\
$\land$ & $\lor$  &.& $(a \lor b) \land c$ & $= a \land c \lor b \land c$\\
$\land$ &.& $\lor$ & $a \land (b \lor c)$ & $= a \land b \lor a \land c$\\
$\lnot$ & $\lor$  & & $\lnot (a \lor b)$ & $= \lnot a \land \lnot b$ \\
$\lnot$ & $\land$ & & $\lnot (a \land b)$ & $= \lnot a \lor \lnot b$ \\
$\lnot$ & $\lnot$ & & $\lnot \lnot a$      & $= a$	\\
\cline{1-3}
\end{tabular}
\caption{Example user-defined signatures (left) and redices (right)}
\LABEL{Example user-defined signatures (left) and redices (right)}
\end{figure}

\begin{figure}
\renewcommand{\arraystretch}{1.0}
\begin{tabular}[t]{l|rr|}
\multicolumn{1}{r}{}
	& \multicolumn{1}{r}{1} 	
	& \multicolumn{1}{r}{2}	\\
\cline{2-3}
$0$     &    & 	\\
$v$     &    & 	\\
$+$     &  1 &  7	\\
$-$     &  1 &  1	\\
$*$     &  7 & 12	\\
$<$     &  1 &  1	\\
$\lor$  & 16 & 21	\\
$\land$ & 25 & 29	\\
$\lnot$ & 32 & 	\\
\cline{2-3}
\end{tabular}
\hfill
\begin{tabular}[t]{r|c@{}c@{}c@{}c@{}c@{}c}
\multicolumn{1}{r}{}
	& \multicolumn{1}{r}{0}
	& \multicolumn{1}{r}{1}
	& \multicolumn{1}{r}{2}
	& \multicolumn{1}{r}{3}
	& \multicolumn{1}{r}{4}
	& \multicolumn{1}{r}{5}	\\
\cline{2-7}
 0: & NIL    &         &         &         &     &	\\
 1: & $0$    & $v$     & $+$     & $-$     & $*$ & NIL	\\
 7: & $0$    & $v$     & $-$     & $*$     & NIL &	\\
12: & $0$    & $v$     & $-$     & NIL     &     &	\\
16: & $<$    & $\lor$  & $\land$ & $\lnot$ & NIL &	\\
21: & $<$    & $\land$ & $\lnot$ & NIL     &     &	\\
25: & $<$    & $\land$ & $\lnot$ & NIL     &     &	\\
29: & $<$    & $\lnot$ & NIL     &         &     &	\\
32: & $<$    & NIL     &         &         &     &	\\
\end{tabular}
\caption{\code{mainIndexTab} (left) and \code{mainArgTab} (right)
resulting from
Fig.~\ref{Example user-defined signatures (left) and redices (right)}}
\LABEL{mainIndexTab (left) and mainArgTab (right) resulting from Fig.}
\end{figure}

In order to exclude nonsensical terms like e.g.\ $3 + (2>1)$
from the build process in Alg.~\ref{Value-pair cached term generation},
we implemented many-sorted signatures for operators.
~
The user can define arbitrarily many sorts, each by either listing its
member values 
(cf.\ module~\ref{valDefTab.c} in Sect.~\ref{Kernel modules overview})
or by naming its \code{print} and \code{scan} routines
(defined in \code{UserVal.c}, cf.\ module~\ref{UserVal.c} in
Sect.\ref{User modules overview}).
~
Each operator is declared having a fixed signature of input sorts and
a result sort 
(module~\ref{opTab.c} in Sect.~\ref{Kernel modules overview}).

In file \code{sorts.c},
we maintain tables \code{mainIndexTab} and \code{mainArgTab}
about which operators
are allowed at a given argument position of a given main operator.
~
This way, in step~\ref{cached term-gen build}
of Alg.~\ref{Term generation from decomposition lists}
we need to consider only terms $t_i$ starting with an operator allowed
at position $i$ of the main operator $f$.
~
Given $w$, $f$, and $i$,
the function \code{NextInhabitedArgOp} in file \code{contTab.c} is used to
enumerate all
operators $f_i$ such that $\D_{w,f_i} \neq \set{}$ and $f_i$ is
allowed at position $i$ of $f$.

In addition to sort information, simple redex information can be
considered in the tables \code{mainIndexTab} and \code{mainArgTab}.
~
For the example redex $x_1 * (x_{21} + x_{22})$ 
from Sect.~\ref{Module redex.c}, we can remove ``$+$'' from the list
of allowed operators at position 2 of ``$*$''.
~
Similarly, we can remove ``$*$'' from the same list if we know that
``$*$'' is associative.
~
Moreover, an uninhabited or unneeded sort
(see module~\ref{opTab.c} in Sect.~\ref{Kernel modules overview})
can be removed from all such lists.

Figure~\ref{Example user-defined signatures (left) and redices (right)}
(left)
shows a user-defined example set of operator signatures and
properties, with ``c'', ``a'', and ``i'' indicating commutativity,
associativity, and idempotency, respectively.
~
Figure~\ref{Example user-defined signatures (left) and redices (right)}
(right)
show a user-defined list of simple redices, with comments or
justifying equations right of the box.
~
Figure~\ref{mainIndexTab (left) and mainArgTab (right) resulting from Fig.}
(left) and (right) shows the layout of the resulting
\code{mainIndexTab} and \code{mainArgTab}, respectively.
~
The former is indexed (vertically) with an operator and (horizontally)
with an argument position, yielding in turn the index where the list
of allowed operators starts in
\code{mainArgTab}.
~
For example, \code{mainIndexTab[*,1] = 7} and \code{mainArgTab[7]}
represents the \code{NIL}-terminated list 
\code{0}, \code{v}, \code{-}, \code{*}.

\newcommand{\inlg}{{\sf K}}

\subsection{Projection-like operators}
\LABEL{Projection-like operators}

In many applications of $E$-generalization, or just of
function synthesis, considering ${\it if}(\cdot,\cdot,\cdot)$
or a similar operator
is indispensable at least from a practical point of view 
(cf.\ the discussion in Exm.~\ref{Interleaved sequence}).
~
In this Section, we formally define the general notion of a 
\notion{projection-like operator}, and
describe implementation optimizations for 
Alg.\ref{Value-pair cached term generation}
related to that operator class.

We call $f: \NF^{m+n} \ra \NF$ a projection-like operator,
if there is a partition of its arguments into (w.l.o.g.)
$\NF^m \times \NF^n$ such that always
$$f(\:,1m{c_\i},\:,1n{d_\i}) 
\in \set{ \:,1n{\vec{d}_\i} } \cup \set{ \undef }$$
and it depends only on $\:,1m{c_\i}$ which argument position is  
``switched  through'' to the output,
that is, if
$$\forall \:,1m{c_\i} \; \exists i \; \forall \:,1n{d_\i}: \;
f(\:,1m{c_\i},\:,1n{d_\i}) = d_i$$
holds.
~
We call $\:,1m{c_\i} \in \NF^m$ the control and $\:,1n{d_\i} \in \NF^n$
the data input to $f$.

Examples of projection-like operators include:\footnote{
	All operator results are $\undef$ for all input value
	combinations not explicitly shown.
}
\begin{itemize}
\item
	Projection
	~
	$\pi^n_i: \NF^0 \times \NF^n \ra \NF$,
	with
	$\pi^n_i(\:,1n{d_\i}) = d_i$;
\item
	Index
	~
	$\idx: \NF \times \NF^n \ra \NF$,
	with
	$\idx(c,\:,1n{d_\i}) = d_c$,
	for $1 \leq c \leq n$;
\item
	If-then-else
	~
	${\it if}: \NF \times \NF^2\ra \NF$,
	with
	${\it if}(\true,d_1,d_2) = d_1$
	and
	${\it if}(\false,d_1,d_2) = d_2$;
\item
	Ifdef
	~
	$?: \NF \times \NF\ra \NF$,
	with
	$?(\true,d) = d$;
\item
	Choice
	~
	$\#^n: \NF^n \times \NF^n \ra \NF$,
	~
	with
	$\#^n (\:,1n{c_\i},\:,1n{d_\i}) = d_i$
	\\
	if $i$ is minimal with
	$\set{\:,1n{c_\i}} \subseteq \set{c_i, \undef}$.
	\\
	The classical angelic choice operator returns $\undef$ if its
	welldefined input values disagree, and the unique welldefined 
	input value, else.
	~
	It can be modeled by $\#^n(\:,1n{x_\i},\:,1n{x_\i})$, 
	taking the same input as both control and data input.
\end{itemize}

Provided all operators are \notion{strict}, i.e.\ return $\undef$
whenever one of their inputs is $\undef$, it possible to write every
term such that never a
projection-like operator occurs below a non-projection-like one.
~
Formally, let $f_i: \NF^{m_i} \times \NF^{n_i} \ra \NF$ 
be projection-like for $i=\:,1l\i$,
and $g: \NF^l \ra \NF$ arbitrary.
~
Each term
$$g( \:,1l{f_\i(\:,1{m_\i}{c_{\i\j}} , \:,1{n_\i}{d_{\i\j}})} )$$
equals a term
$$f( \:,1l{\:,1{m_\i}{c_{\i\j}}}, \:,1{...}{g_\i(...)} )$$
for a suitable projection-like operator
$f: \NF^{\:+1l{m_\i}} \times \NF^{...} \ra \NF$.
~
Therefore, we don't lose completeness of 
Alg.~\ref{Value-pair cached term generation}
if we modify it such that projection-like operators appear only at the
top of built terms.
~
However, moving projection-like operators to the top
may result in exponential term growth.

Since our implementation optimizations don't make much sense when the
$E$-generalization of only $2$ terms is computed, we assume in the
rest of this section that an arbitrary number $\inlg$ of terms
$\:,1\inlg{t_\i}$ is to be generalized simultaneously.
~
The algorithms from
Sect.~\ref{An improved algorithm to compute E-generalizations}
can be extended in a straight-forward way.
~
We call a subset of $\set{ \:,1\inlg\i }$ an index set.

\begin{figure}
\begin{center}

\definecolor{ceLat}	{rgb}{0.90,0.90,0.90}	
\definecolor{ceUpd}	{rgb}{0.00,0.99,0.50}	
\definecolor{cePrn}	{rgb}{0.99,0.30,0.50}	
\definecolor{cePrd}	{rgb}{0.99,0.60,0.80}	

\definecolor{cnUpd}	{rgb}{0.00,0.60,0.00}	
\definecolor{cnPrn}	{rgb}{0.60,0.00,0.00}	
\definecolor{cnPrd}	{rgb}{0.99,0.50,0.70}	

\setlength{\unitlength}{0.75mm}


\newcommand{\node}[4]{
	\put(#1,#2){\circle*{1}}%
	\put(#1,#2){\put(1,0){\makebox(0,0)[l]{$#3$}}}%
	\put(#1,#2){\put(-1,0){\makebox(0,0)[r]{$\scriptstyle\sf #4$}}}%
}

\begin{picture}(200,150)

\thicklines\textcolor{cePrd}{\put(90.000,0.000){\line(-3,1){90.000}}}
\thinlines \textcolor{ceLat}{\put(90.000,0.000){\line(-3,2){45.000}}}
\thinlines \textcolor{ceLat}{\put(90.000,0.000){\line(0,1){30.000}}}
\thinlines \textcolor{ceLat}{\put(90.000,0.000){\line(3,2){45.000}}}
\thinlines \textcolor{ceLat}{\put(90.000,0.000){\line(3,1){90.000}}}
\thicklines\textcolor{cnPrd}{\node{90}{0}{\set{}}{0}}

\thicklines\textcolor{cePrd}{\put(0.000,30.000){\line(0,1){30.000}}}
\thinlines \textcolor{ceLat}{\put(0.000,30.000){\line(2,3){20.000}}}
\thinlines \textcolor{ceLat}{\put(0.000,30.000){\line(4,3){40.000}}}
\thinlines \textcolor{ceLat}{\put(0.000,30.000){\line(2,1){60.000}}}
\thicklines\textcolor{cePrd}{\put(45.000,30.000){\line(-3,2){45.000}}}
\thinlines \textcolor{ceLat}{\put(45.000,30.000){\line(6,5){30.000}}}
 \thinlines \textcolor{ceLat}{\put(75.000,55.000){\line(1,1){5.000}}}
\thinlines \textcolor{ceLat}{\put(45.000,30.000){\line(2,1){30.000}}}
 \thinlines \textcolor{ceLat}{\put(75.000,45.000){\line(5,3){25.000}}}
\thinlines \textcolor{ceLat}{\put(45.000,30.000){\line(5,2){75.000}}}
\thicklines\textcolor{cePrd}{\put(90.000,30.000){\line(-2,1){20.000}}}
 \thicklines\textcolor{cePrd}{\put(70.000,40.000){\line(-5,2){50.000}}}
\thinlines \textcolor{ceLat}{\put(90.000,30.000){\line(-1,3){10.000}}}
\thinlines \textcolor{ceLat}{\put(90.000,30.000){\line(5,3){50.000}}}
\thinlines \textcolor{ceLat}{\put(90.000,30.000){\line(5,2){50.000}}}
 \thinlines \textcolor{ceLat}{\put(140.000,50.000){\line(2,1){20.000}}}
\thicklines\textcolor{ceUpd}{\put(135.000,30.000){\line(-3,1){75.000}}}
 \thicklines\textcolor{ceUpd}{\put(60.000,55.000){\line(-4,1){20.000}}}
\thinlines \textcolor{ceLat}{\put(135.000,30.000){\line(-1,1){5.000}}}
 \thinlines \textcolor{ceLat}{\put(130.000,35.000){\line(-6,5){30.000}}}
\thinlines \textcolor{ceLat}{\put(135.000,30.000){\line(1,6){5.000}}}
\thinlines \textcolor{ceLat}{\put(135.000,30.000){\line(3,2){45.000}}}
\thicklines\textcolor{cePrd}{\put(180.000,30.000){\line(-4,1){120.000}}}
\thinlines \textcolor{ceLat}{\put(180.000,30.000){\line(-2,1){60.000}}}
\thinlines \textcolor{ceLat}{\put(180.000,30.000){\line(-2,3){20.000}}}
\thinlines \textcolor{ceLat}{\put(180.000,30.000){\line(0,1){30.000}}}
\textcolor{cnPrd}{\node{  0}{30}{1}{2}}
\textcolor{cnPrd}{\node{ 45}{30}{2}{1}}
\textcolor{cnPrd}{\node{ 90}{30}{3}{2}}
\textcolor{cnUpd}{\node{135}{30}{4}{\infty}}
\textcolor{cnPrd}{\node{180}{30}{5}{1}}

\thicklines\textcolor{cePrd}{\put(0.000,60.000){\line(0,1){30.000}}}
\thinlines \textcolor{ceLat}{\put(0.000,60.000){\line(2,3){20.000}}}
\thinlines \textcolor{ceLat}{\put(0.000,60.000){\line(4,3){40.000}}}
\thicklines\textcolor{cePrd}{\put(20.000,60.000){\line(-2,3){20.000}}}
\thinlines \textcolor{ceLat}{\put(20.000,60.000){\line(4,3){40.000}}}
\thinlines \textcolor{ceLat}{\put(20.000,60.000){\line(2,1){60.000}}}
\thicklines\textcolor{ceUpd}{\put(40.000,60.000){\line(-2,3){20.000}}}
\thinlines \textcolor{ceLat}{\put(40.000,60.000){\line(2,3){20.000}}}
\thinlines \textcolor{ceLat}{\put(40.000,60.000){\line(2,1){60.000}}}
\thicklines\textcolor{cePrd}{\put(60.000,60.000){\line(-2,3){20.000}}}
\thinlines \textcolor{ceLat}{\put(60.000,60.000){\line(2,3){20.000}}}
\thinlines \textcolor{ceLat}{\put(60.000,60.000){\line(4,3){40.000}}}
\thicklines\textcolor{cePrd}{\put(80.000,60.000){\line(-5,2){50.000}}}
 \thicklines\textcolor{cePrd}{\put(30.000,80.000){\line(-3,1){30.000}}}
\thinlines \textcolor{ceLat}{\put(80.000,60.000){\line(4,3){40.000}}}
\thinlines \textcolor{ceLat}{\put(80.000,60.000){\line(2,1){60.000}}}
\thicklines\textcolor{ceUpd}{\put(100.000,60.000){\line(-5,2){50.000}}}
 \thicklines\textcolor{ceUpd}{\put(50.000,80.000){\line(-3,1){30.000}}}
\thinlines \textcolor{ceLat}{\put(100.000,60.000){\line(2,3){20.000}}}
\thinlines \textcolor{ceLat}{\put(100.000,60.000){\line(2,1){60.000}}}
\thicklines\textcolor{cePrd}{\put(120.000,60.000){\line(-5,2){50.000}}}
 \thicklines\textcolor{cePrd}{\put(70.000,80.000){\line(-3,1){30.000}}}
\thinlines \textcolor{ceLat}{\put(120.000,60.000){\line(2,3){20.000}}}
\thinlines \textcolor{ceLat}{\put(120.000,60.000){\line(4,3){40.000}}}
\thicklines\textcolor{ceUpd}{\put(140.000,60.000){\line(-5,2){50.000}}}
 \thicklines\textcolor{ceUpd}{\put(90.000,80.000){\line(-3,1){30.000}}}
\thinlines \textcolor{ceLat}{\put(140.000,60.000){\line(-2,3){20.000}}}
\thinlines \textcolor{ceLat}{\put(140.000,60.000){\line(4,3){40.000}}}
\thicklines\textcolor{cePrd}{\put(160.000,60.000){\line(-5,2){50.000}}}
 \thicklines\textcolor{cePrd}{\put(110.000,80.000){\line(-3,1){30.000}}}
\thinlines \textcolor{ceLat}{\put(160.000,60.000){\line(-2,3){20.000}}}
\thinlines \textcolor{ceLat}{\put(160.000,60.000){\line(2,3){20.000}}}
\thicklines\textcolor{ceUpd}{\put(180.000,60.000){\line(-5,2){50.000}}}
 \thicklines\textcolor{ceUpd}{\put(130.000,80.000){\line(-3,1){30.000}}}
\thinlines \textcolor{ceLat}{\put(180.000,60.000){\line(-2,3){20.000}}}
\thinlines \textcolor{ceLat}{\put(180.000,60.000){\line(0,1){30.000}}}
\textcolor{cnPrd}{\node{  0}{60}{12}{3}}
\textcolor{cnPrd}{\node{ 20}{60}{13}{3}}
\textcolor{cnUpd}{\node{ 40}{60}{14}{\infty}}
\textcolor{cnPrd}{\node{ 60}{60}{15}{2}}
\textcolor{cnPrd}{\node{ 80}{60}{23}{3}}
\textcolor{cnUpd}{\node{100}{60}{24}{\infty}}
\textcolor{cnPrd}{\node{120}{60}{25}{2}}
\textcolor{cnUpd}{\node{140}{60}{34}{\infty}}
\textcolor{cnPrd}{\node{160}{60}{35}{2}}
\textcolor{cnUpd}{\node{180}{60}{45}{\infty}}

\thicklines\textcolor{cePrn}{\put(0.000,90.000){\line(0,1){30.000}}}
\thinlines \textcolor{ceLat}{\put(0.000,90.000){\line(3,2){45.000}}}
\thicklines\textcolor{ceUpd}{\put(20.000,90.000){\line(-2,3){20.000}}}
\thinlines \textcolor{ceLat}{\put(20.000,90.000){\line(5,2){50.000}}}
 \thinlines \textcolor{ceLat}{\put(70.000,110.000){\line(2,1){20.000}}}
\thicklines\textcolor{cePrd}{\put(40.000,90.000){\line(1,6){5.000}}}
\thinlines \textcolor{ceLat}{\put(40.000,90.000){\line(5,3){50.000}}}
\thicklines\textcolor{ceUpd}{\put(60.000,90.000){\line(-2,1){60.000}}}
\thinlines \textcolor{ceLat}{\put(60.000,90.000){\line(5,2){75.000}}}
\thicklines\textcolor{cePrd}{\put(80.000,90.000){\line(-1,1){5.000}}}
 \thicklines\textcolor{cePrd}{\put(75.000,95.000){\line(-6,5){30.000}}}
\thinlines \textcolor{ceLat}{\put(80.000,90.000){\line(2,1){30.000}}}
 \thinlines \textcolor{ceLat}{\put(110.000,105.000){\line(5,3){25.000}}}
\thicklines\textcolor{ceUpd}{\put(100.000,90.000){\line(-1,3){10.000}}}
\thinlines \textcolor{ceLat}{\put(100.000,90.000){\line(6,5){30.000}}}
 \thinlines \textcolor{ceLat}{\put(130.000,115.000){\line(1,1){5.000}}}
\thicklines\textcolor{ceUpd}{\put(120.000,90.000){\line(-4,1){120.000}}}
\thinlines \textcolor{ceLat}{\put(120.000,90.000){\line(2,1){60.000}}}
\thicklines\textcolor{cePrd}{\put(140.000,90.000){\line(-3,1){75.000}}}
 \thicklines\textcolor{cePrd}{\put(65.000,115.000){\line(-4,1){20.000}}}
\thinlines \textcolor{ceLat}{\put(140.000,90.000){\line(4,3){40.000}}}
\thicklines\textcolor{ceUpd}{\put(160.000,90.000){\line(-2,1){20.000}}}
 \thicklines\textcolor{ceUpd}{\put(140.000,100.000){\line(-5,2){50.000}}}
\thinlines \textcolor{ceLat}{\put(160.000,90.000){\line(2,3){20.000}}}
\thicklines\textcolor{ceUpd}{\put(180.000,90.000){\line(-3,2){45.000}}}
\thinlines \textcolor{ceLat}{\put(180.000,90.000){\line(0,1){30.000}}}
\textcolor{cnPrn}{\node{  0}{90}{123}{4}}
\textcolor{cnUpd}{\node{ 20}{90}{124}{\infty}}
\textcolor{cnPrd}{\node{ 40}{90}{125}{3}}
\textcolor{cnUpd}{\node{ 60}{90}{134}{\infty}}
\textcolor{cnPrd}{\node{ 80}{90}{135}{3}}
\textcolor{cnUpd}{\node{100}{90}{145}{\infty}}
\textcolor{cnUpd}{\node{120}{90}{234}{\infty}}
\textcolor{cnPrd}{\node{140}{90}{235}{3}}
\textcolor{cnUpd}{\node{160}{90}{245}{\infty}}
\textcolor{cnUpd}{\node{180}{90}{345}{\infty}}

\thicklines\textcolor{ceUpd}{\put(0.000,120.000){\line(3,1){90.000}}}
\thicklines\textcolor{cePrn}{\put(45.000,120.000){\line(3,2){45.000}}}
\thicklines\textcolor{ceUpd}{\put(90.000,120.000){\line(0,1){30.000}}}
\thicklines\textcolor{ceUpd}{\put(135.000,120.000){\line(-3,2){45.000}}}
\thicklines\textcolor{ceUpd}{\put(180.000,120.000){\line(-3,1){90.000}}}
\textcolor{cnUpd}{\node{  0}{120}{1234}{\infty}}%
\textcolor{cnPrn}{\node{ 45}{120}{1235}{4}}%
\textcolor{cnUpd}{\node{ 90}{120}{1245}{\infty}}%
\textcolor{cnUpd}{\node{135}{120}{1345}{\infty}}%
\textcolor{cnUpd}{\node{180}{120}{2345}{\infty}}%

\textcolor{cnUpd}{\node{ 90}{150}{12345}{\infty}}%

\end{picture}

\caption{Example update of data term lattice}
\LABEL{Example update of data term lattice}
\end{center}
\end{figure}
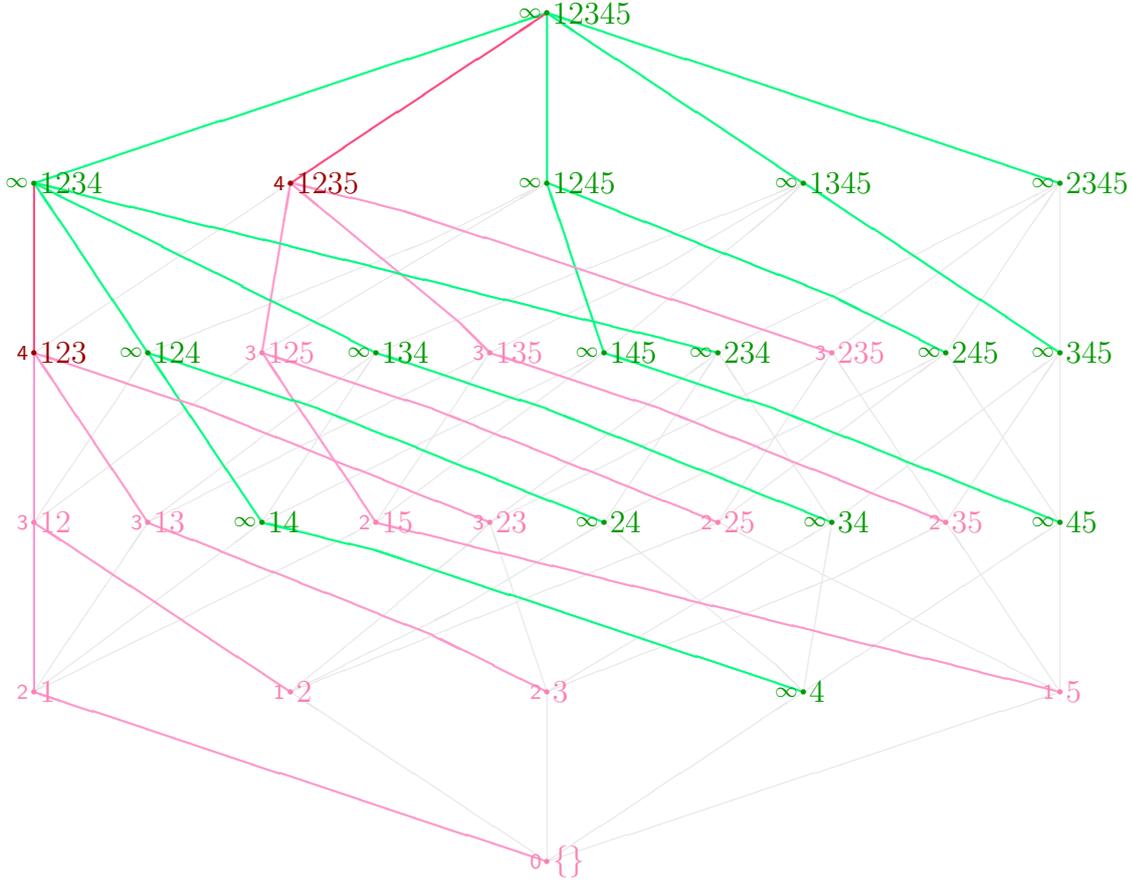

\subsubsection{The data term lattice}
\LABEL{The data term lattice}

We maintain a lattice of all subsets of $\set{ \:,1\inlg\i }$,
i.e.\ all index sets.
~
For every such subset, the lattice holds a ``data''
term of minimal weight found so far
that matches the goal tuple at least at that subset.
~
Whenever a new term is entered into $\D$, its index positions matching
the goal tuple are determined, and the data term lattice is updated
accordingly.
~
The lattice satisfies the invariant that the weight of a node is
larger or equal than that of each of its subset nodes.
~
In particular, its bottom node always has the least weight at all
in $\W$.
~
When no term has yet been found for some index set, its weight is set to
$\infty$.

Figure~\ref{Example update of data term lattice}
shows an example update operation of the data term lattice for
$\inlg=5$,
abbreviating e.g.\ $\set{ 1,2,3,4,5 }$ as $12345$.
~
To the right and left
of each node, the set of indices matching the goal tuple
and the minimal weight found so far
is shown, respectively.
~
Figure~\ref{Example update of data term lattice}
assumes a term of weight 5 matching the goal tuple at position
$\set{ 1,2,3,4,5 }$ has been found recently, and is about to be
entered into the lattice.
~
To this end, all green nodes are to be updated to weight 5,
while red nodes are left unchanged.
~
Updates are attempted along the spanning tree of the sublattice below
$\set{ 1,2,3,4,5 }$; it is obtained by connecting each index set
(e.g.\ $\set{ 1,2,4 }$)
to its leftmost, i.e.\ lexicographically least,
immediate superset (e.g.\ $\set{ 1,2,3,4 }$).
~
Observe that the structure of the spanning tree depends on its topmost
node.
~
For example, the lattice for $\set{ 1,2,3,4 }$
is a sublattice of that for $\set{ 1,2,3,4,5 }$, while
the spanning tree for $\set{ 1,2,3,4 }$
is not a subtree of that for $\set{ 1,2,3,4,5 }$.

While the data term lattice is independent of the particular kind of
projection-like operators in use, the control term data structure
does depend on the latter.
~
We describe the structures and algorithms for the \code{C}
\code{switch} like $\idx$ operator family\footnote{
	The $n+1$-ary operation $\idx_n(c,\:,0{n-1}{d_\i})$
	is defined to yield $d_c$ if $0 \leq c < n$, and $\undef$, else,
	similar to an array indexing expression
	\code{d[c]} in imperative programming languages.
}
in the following;
the usual ternary ${\it if}(\cdot,\cdot,\cdot)$ and some variants can be
handled in a similar way.

\subsubsection{The $\idx$ control term lattice}

\begin{figure}
\begin{center}
\includegraphics{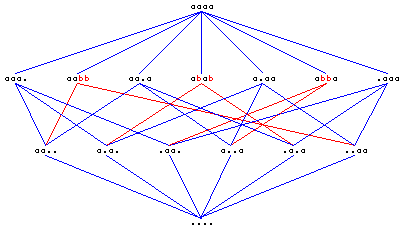}
\caption{Lattice of equivalence relations for $\inlg=4$}
\LABEL{Lattice of equivalence relations for inlg=4}
\end{center}
\end{figure}

We maintain a lattice of all equivalence relations on the set of
indices of value tuples.
~
As an example, Fig.~\ref{Lattice of equivalence relations for inlg=4}
shows the lattice for $\inlg=4$, with ``a'' and ``b'' in an
equivalence relation's node denoting the
partition an element is mapped to by that relation, and ``.''
indicating mapping to a singleton partition; e.g.\ ``aa..'' denotes
the partition
$\set{ a\.=\set{1,2}, \set{3}, \set{4} }$,
corresponding to the equivalence relation 
$\set{ \tpl{1,1}, \tpl{1,2}, \tpl{2,1}, \tpl{2,2}, \tpl{3,3},
\tpl{4,4} }$.

For every such equivalence relation $R$, its lattice node
holds a ``control'' term $t$
of minimal weight found so
far such that $\nf(t \sigma_i) = \nf(t \sigma_j)$
iff $i \mathrel{R} j$.
~
For each $R$,
we also maintain the count of partitions of $R$ for which data terms exist
in the data lattice.
~
When a data term exists for each partition, and a control term exists,
we can build a solution term by composing them appropriately with an
$\idx$ operator.

For example,
from a control term $t_c$ evaluating to $08101$,
and data terms $t_0$, $t_1$, and $t_8$ matching the goal vector at
index $1,4$, at index $3,5$, and at index $2$, respectively,
we can build the term $\idx_9(t_c,t_0,t_1,0,0,0,0,0,0,t_8)$,
which matches the
goal vector at index $1,2,3,4,5$ by construction.
~
The zero arguments of $\idx_9$ are used for padding purposes
only, i.e.\ to get $t_8$ into the right place.

Whenever a new control\footnote{
	A term $t$ is a potential control term if it has
	an appropriate sort
	(e.g.\ \code{int})
	and $0 \leq t \sigma_i < A-1$ for all $i$, where $A$ denotes
	the maximal arity of admitted $\idx$ operators.
}
term is entered into $\D$,
and its node in the control lattice is still empty,
we store it there, and initialize its partition count.
~
An earlier term need never be replaced by a later one, since terms
appear in order of increasing weight.

Whenever a new data term is entered into $\D$,
we update the data term sublattice below it as described above
in Sect.~\ref{The data term lattice}.
~
For each updated sublattice node, corresponding to a set $S$ of
indices,
we increment the partition count in the control
term lattice for all equivalences having $S$ as one of its partitions.
~
These equivalences are found as refinements of the relation
$i \mathrel{R} j \Lra (i \in S \Lra j \in S)$.

\subsection{Module \code{assoc.c}}
\LABEL{Module assoc.c}

This module
implements a lookup mechanism for indices into \code{contTab} on top of
\code{bbt.c} (module~\ref{bbt.c} in Sect.~\ref{Kernel modules overview}).
~
For each user-sort, a hash table is allocated, the size of which is
always some power $c^n$ of the sort's cardinality $c$.
Each hash entry then points to a balanced binary tree.

Figure~\ref{Example hash tables in assoc.c fig} shows an example,\footnote{
	Our implementation includes an own undefined value $\undef_s$
	into every sort $s$.
	~
	For sake of simplicity we ignore them in this example.
}
assuming a sort \code{bool} with values
$\false$ and $\true$,
a sort \code{int} with values \code{0} to \code{9},
operators
$\code{0}, \code{1}, \code{2}, \code{3} : \ra \code{int}$,
$\code{+}, \code{*} : \code{int}, \code{int} \ra \code{int}$,
$\code{<} : \code{int}, \code{int} \ra \code{bool}$,
and variables
$\code{x}, \code{y} : \ra \code{int}$
with $\code{x} \simeq \tpl{\code{5},\code{2},\code{1}}$
and
$\code{y} \simeq \tpl{\code{5},\code{9},\code{2}}$.\footnote{
	i.e.\ with $\inlg = 3$,
	$\sigma_1 = \subst{ x \mapsto 5, y \mapsto 5}$,
	$\sigma_2 = \subst{ x \mapsto 2, y \mapsto 9}$, and
	$\sigma_3 = \subst{ x \mapsto 1, y \mapsto 2}$
}
~
We have $c = 2$, $n = 3$ for \code{bool}, and $c = 10$, $n = 1$ for
\code{int}.
~
The balanced binary tree for the \code{int} entry \code{5} represents
the term set $\set{2+3, x, y}$ of all terms evaluating to some
$\tpl{5,\cdot,\cdot}$ that have been found so far.

The lexicographic comparison of value vectors within a binary tree need
not consider the first $n$ vector components, as they always agree.
~
Therefore, we store the value of $n$ in to field \code{cmpStart} of a
\code{struct \_userSort} 
(module~\ref{opTab.c} in Sect.~\ref{Kernel modules overview}).

If, for some sort,
$c^n$ is sufficiently small, all possible value vectors fit into the
hash table, like for sort \code{bool} in the above example.
~
In this case, only trivial balanced binary trees occur as hash
entries.
~
Moreover, when each hash entry is filled, we know that a minimal term
has already been generated for each value vector; and we need not
build any more terms of the sort.
~
For this purpose, we use the field \code{hashEmptyCnt}
of a \code{struct \_userSort} and the flag
\code{usfSaturated}.
~
The latter is checked by \code{computeHeapSequence} in \code{compute.c}.

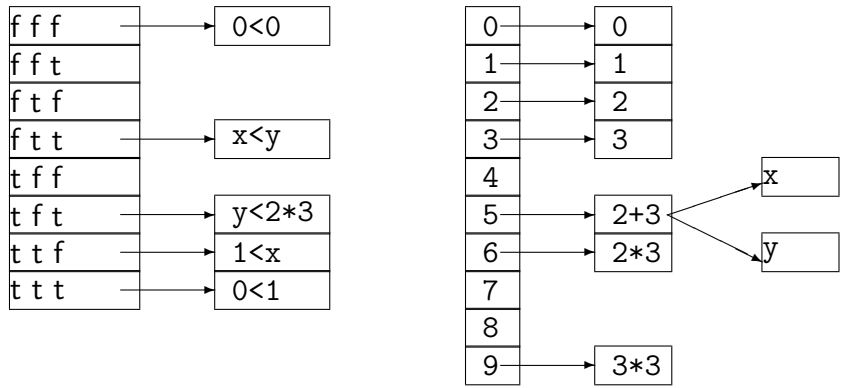
\begin{figure}
\begin{center}
\begin{picture}(110,50)
 \put(0.000,45.000){\framebox(17.000,5.000)[l]{\false\ \false\ \false}}
 \put(0.000,40.000){\framebox(17.000,5.000)[l]{\false\ \false\ \true }}
 \put(0.000,35.000){\framebox(17.000,5.000)[l]{\false\ \true \ \false}}
 \put(0.000,30.000){\framebox(17.000,5.000)[l]{\false\ \true \ \true }}
 \put(0.000,25.000){\framebox(17.000,5.000)[l]{\true \ \false\ \false}}
 \put(0.000,20.000){\framebox(17.000,5.000)[l]{\true \ \false\ \true }}
 \put(0.000,15.000){\framebox(17.000,5.000)[l]{\true \ \true \ \false}}
 \put(0.000,10.000){\framebox(17.000,5.000)[l]{\true \ \true \ \true }}
 \put(14.500,47.500){\vector(1,0){12.500}}
 \put(14.500,32.500){\vector(1,0){12.500}}
 \put(14.500,22.500){\vector(1,0){12.500}}
 \put(14.500,17.500){\vector(1,0){12.500}}
 \put(14.500,12.500){\vector(1,0){12.500}}
 \put(27.000,45.000){\framebox(15.000,5.000)[l]{\code{ 0<0}}}
 \put(27.000,30.000){\framebox(15.000,5.000)[l]{\code{ x<y}}}
 \put(27.000,20.000){\framebox(15.000,5.000)[l]{\code{ y<2*3}}}
 \put(27.000,15.000){\framebox(15.000,5.000)[l]{\code{ 1<x}}}
 \put(27.000,10.000){\framebox(15.000,5.000)[l]{\code{ 0<1}}}
 \put(60.000,45.000){\framebox(7.000,5.000)[l]{\code{ 0}}}
 \put(60.000,40.000){\framebox(7.000,5.000)[l]{\code{ 1}}}
 \put(60.000,35.000){\framebox(7.000,5.000)[l]{\code{ 2}}}
 \put(60.000,30.000){\framebox(7.000,5.000)[l]{\code{ 3}}}
 \put(60.000,25.000){\framebox(7.000,5.000)[l]{\code{ 4}}}
 \put(60.000,20.000){\framebox(7.000,5.000)[l]{\code{ 5}}}
 \put(60.000,15.000){\framebox(7.000,5.000)[l]{\code{ 6}}}
 \put(60.000,10.000){\framebox(7.000,5.000)[l]{\code{ 7}}}
 \put(60.000,5.000){\framebox(7.000,5.000)[l]{\code{ 8}}}
 \put(60.000,0.000){\framebox(7.000,5.000)[l]{\code{ 9}}}
 \put(64.500,47.500){\vector(1,0){12.500}}
 \put(64.500,42.500){\vector(1,0){12.500}}
 \put(64.500,37.500){\vector(1,0){12.500}}
 \put(64.500,32.500){\vector(1,0){12.500}}
 \put(64.500,22.500){\vector(1,0){12.500}}
 \put(64.500,17.500){\vector(1,0){12.500}}
 \put(64.500,2.500){\vector(1,0){12.500}}
 \put(77.000,45.000){\framebox(10.000,5.000)[l]{\code{ 0}}}
 \put(77.000,40.000){\framebox(10.000,5.000)[l]{\code{ 1}}}
 \put(77.000,35.000){\framebox(10.000,5.000)[l]{\code{ 2}}}
 \put(77.000,30.000){\framebox(10.000,5.000)[l]{\code{ 3}}}
 \put(77.000,20.000){\framebox(10.000,5.000)[l]{\code{ 2+3}}}
 \put(77.000,15.000){\framebox(10.000,5.000)[l]{\code{ 2*3}}}
 \put(77.000,0.000){\framebox(10.000,5.000)[l]{\code{ 3*3}}}
 \put(86.500,22.500){\vector(3,1){12.500}}
 \put(86.500,22.500){\vector(2,-1){12.500}}
 \put(99.000,25.000){\framebox(10.000,5.000)[l]{\code{x}}}
 \put(99.000,15.000){\framebox(10.000,5.000)[l]{\code{y}}}
\end{picture}
\caption{Example hash tables in \code{assoc.c} 
	(Sect.~\ref{Module assoc.c})}
\LABEL{Example hash tables in assoc.c fig}
\end{center}
\end{figure}

\section{Run time statistics}
\LABEL{Run time statistics}

In this section, we give some statistical figures about a typical run
of our implementation.
~
In order to compute a law term for the IST'70-ZR test ``A1 102'',\footnote{
	This test asks for a law of
	$1; 3, 6, 8, 16, 18, 36$;
	e.g.\ the term
	${\it if}(v_p \% 2 = 0, v_1+v_1, v_1+2)$
	is a ``correct'' solution.
}
we set $\inlg=6$ and
search a term for the value vector $\tpl{3, 6, 8, 16, 18, 36}$,
given the variables $v_1 = \tpl{ 1, 3, 6, 8, 16, 18}$
and $v_p = \tpl{0, 1, 2, 3, 4, 5}$; 
admitted operators were $+, -, *, /, /\!/, \%, \idx_2, \idx_3, \ldots$, 
all weight functions were such that weight and size\footnote{
	See Exm.~\ref{Weight function examples}.
}
of a term agreed.
~
Before the first second elapsed,
during the computation of terms of weight 5,
the terms
$v_p \% 2$, $v_1*2$, and $v_1+2$ became available\footnote{
	Due to sort inclusion functions that are omitted here for
	simplicity, the terms have a size of 4, 5, and 5, respectively.
}
and the term
$\idx_2(v_p \% 2, v_1*2, v_1+2)$
was built and found to be a solution.
~
However, the search continued since this solution was not known to be
of minimal weight, while a minimal solution was required by a
command-line option.
~
After 210 seconds of user time, the memory was exhausted during build
of weight level 12, and the implementation aborted, returning the
above term as the best available solution.

\begin{figure}
\begin{center}
\includegraphics[width=\textwidth]{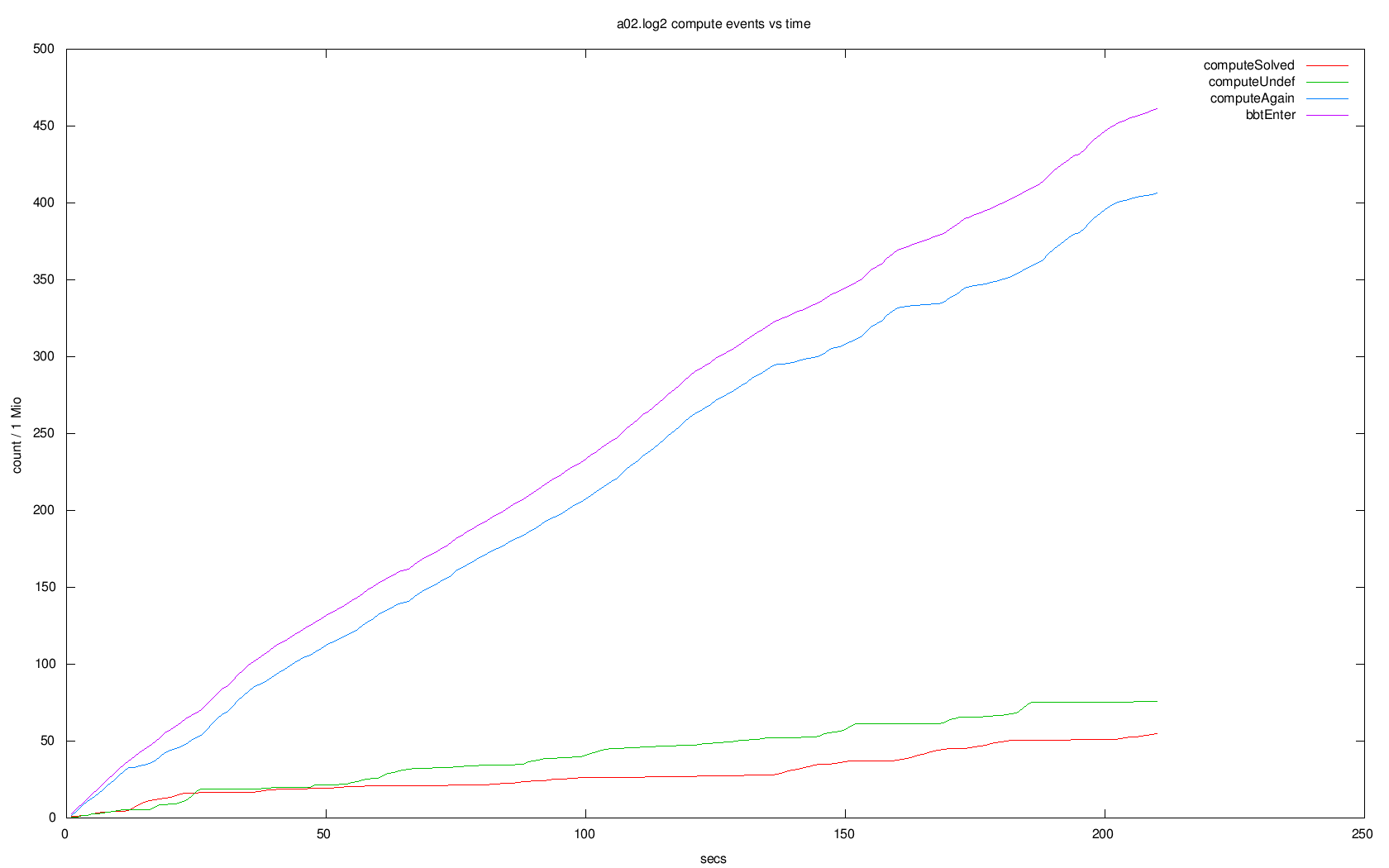}
\caption{Compute event statistics}
\label{Compute event statistics}
\end{center}
\end{figure}

Figure~\ref{Compute event statistics}
shows the number of terms for which the value vectors were computed
and turned out to be new (``computeSolved''), already known
(``computeAgain''), undefined (``computeUndef''), and for which
$\phi$ was looked up 
(``bbtEnter'', cf.\ Fig.~\ref{Example state in Alg. (red) and (blue)});
the latter equals the sum of ``computeSolved'' and ``computeAgain''.

\begin{figure}
\begin{center}
\includegraphics[width=\textwidth]{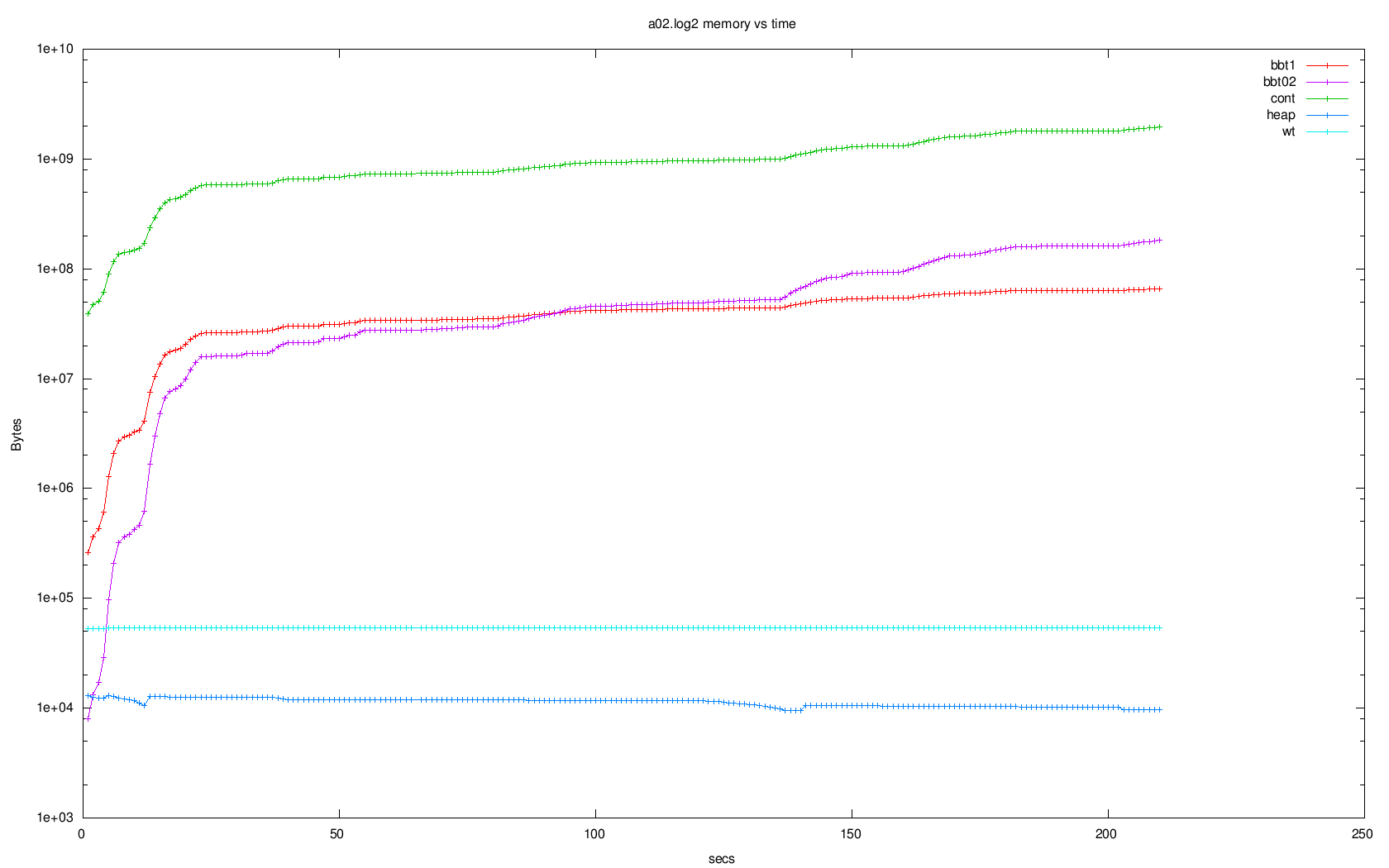}
\caption{Memory statistics}
\label{Memory statistics}
\end{center}
\end{figure}

Figure~\ref{Memory statistics} shows the memory consumption 
\begin{itemize}
\item for the balanced binary tree implementing $\phi$
	(``bbt1'' and ``bbt02'', corresponding to nodes with 1 and 2
	children, respectively),
\item for $\D$ (``cont''),
\item for $\F$ (``heap''), and
\item for the table holding weight terms 
	(``wt'', pointed-to by ``heap'' entries).
\end{itemize}

\begin{figure}
\begin{center}
\includegraphics[width=\textwidth]{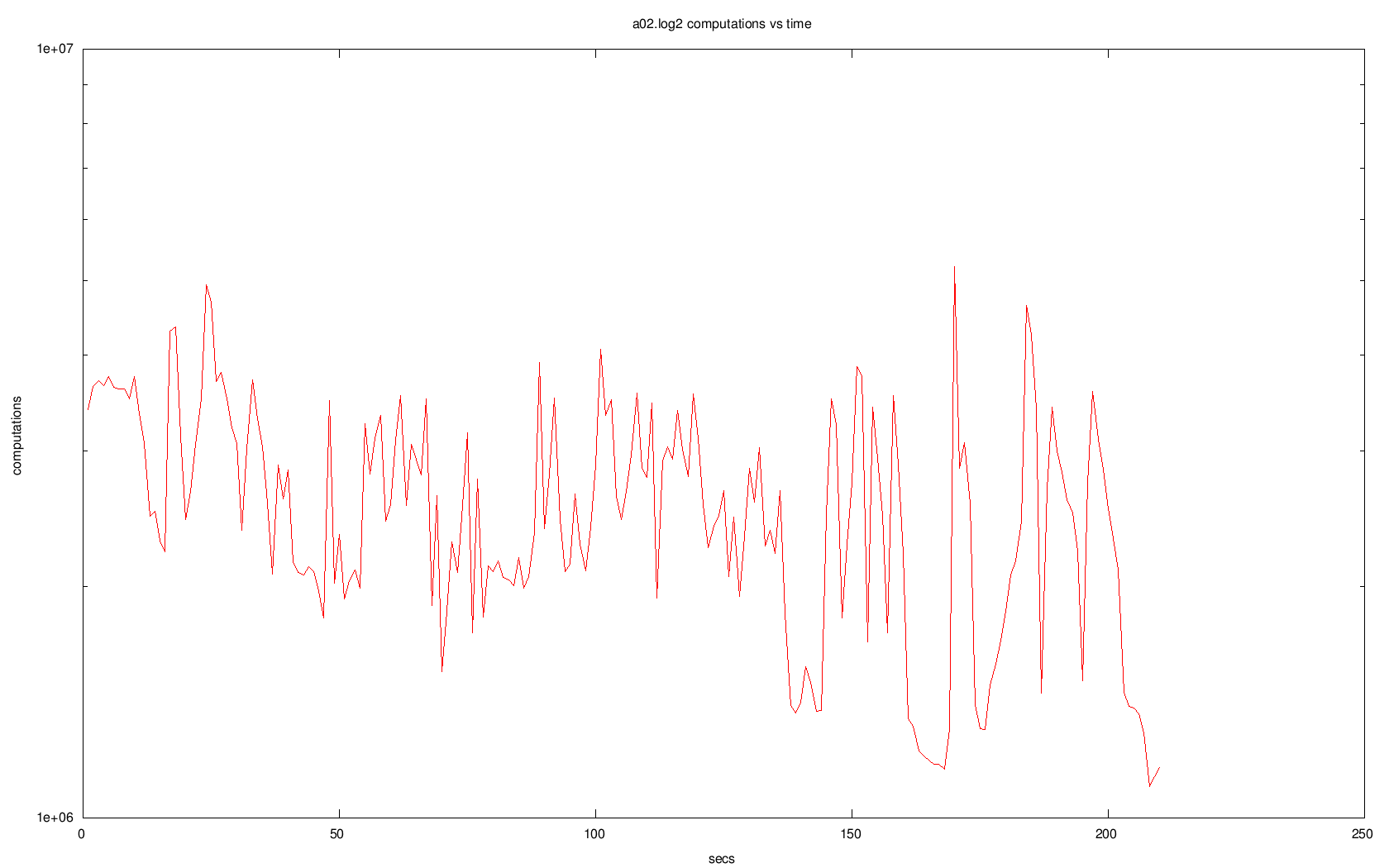}
\caption{Value vectors statistics}
\label{Value vectors statistics}
\end{center}
\end{figure}

\begin{figure}
\begin{center}
\includegraphics[width=\textwidth]{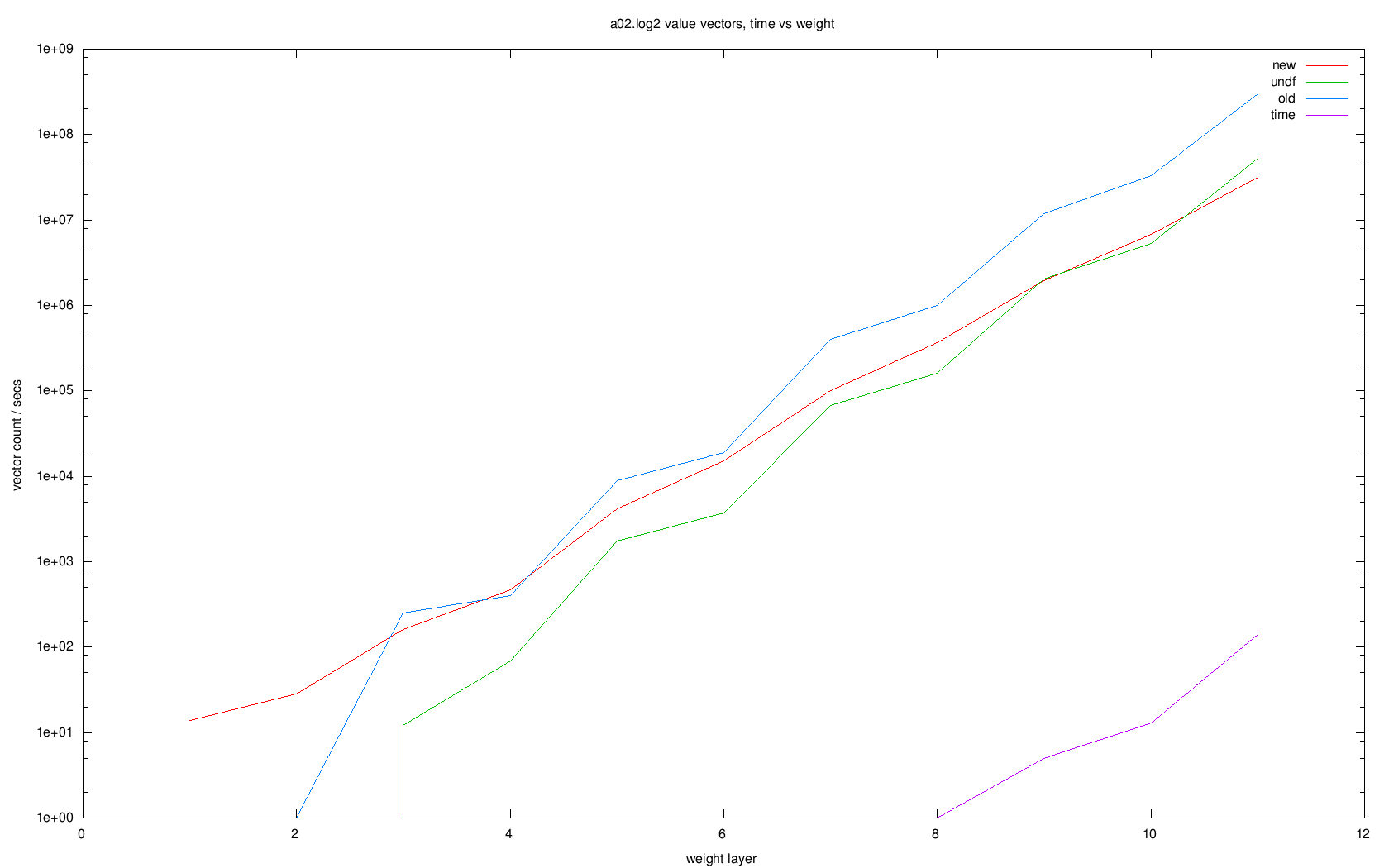}
\caption{Weight statistics}
\label{Weight statistics}
\end{center}
\end{figure}

Figure~\ref{Value vectors statistics} shows the number of computed terms
for each second of elapsed time.
~
Finally,
Fig.~\ref{Weight statistics} shows for each weight
the number of new, undefined, and old terms of that weight, and the
total time spent with their computation.

\clearpage


\clearpage

\bibliographystyle{alpha}
\bibliography{lit}

\end{document}